\documentclass[prd,nofootinbib,showpacs,superscriptaddress,preprint]{revtex4}
\usepackage[T1]{fontenc}
\usepackage{amsmath,amssymb}
\usepackage{epsfig}
\usepackage{graphicx}
\usepackage[usenames,dvipsnames]{color}
\usepackage{slashed}
\usepackage[colorlinks,citecolor=blue]{hyperref}
\usepackage{pdfpages}
\usepackage{color}
\begin{document}
\title{Common Origin of Non-zero $\theta_{13}$ and Dark Matter in an $S_4$ Flavour Symmetric Model with Inverse Seesaw}

\author{Ananya Mukherjee}
\email{ananyam@tezu.ernet.in}
\affiliation{Department of Physics, Tezpur University, Tezpur - 784028, India}

\author{Debasish Borah}
\email{dborah@iitg.ernet.in}
\affiliation{Department of Physics, Indian Institute of Technology Guwahati, Assam 781039, India}
\author{Mrinal Kumar Das}
\email{mkdas@tezu.ernet.in}
\affiliation{Department of Physics, Tezpur University, Tezpur - 784028, India}

\begin{abstract}
We study an inverse seesaw model of neutrino mass within the framework of $S_4$ flavour symmetry from the requirement of generating non-zero reactor mixing angle $\theta_{13}$ along with correct dark matter relic abundance. The leading order $S_4$ model gives rise to tri-bimaximal type leptonic mixing resulting in $\theta_{13}=0$. Non-zero $\theta_{13}$ is generated at one loop level by extending the model with additional scalar and fermion fields which take part in the loop correction. The particles going inside the loop are odd under an in-built $Z^{\text{Dark}}_2$ symmetry such that the lightest $Z^{\text{Dark}}_2$ odd particle can be a dark matter candidate. Correct neutrino and dark matter phenomenology can be achieved for such one loop corrections either to the light neutrino mass matrix or to the charged lepton mass matrix although the latter case is found to be more predictive. The predictions for neutrinoless double beta decay is also discussed and inverted hierarchy in the charged lepton correction case is found to be disfavoured by the latest KamLAND-Zen data.
\end{abstract}
\maketitle


\section{Introduction}
\label{sec:level1} 
The Standard Model (SM) of particle physics surmises on the minimal choice that a single Higgs doublet provides masses to all particles. Some questions however remain unanswered, including the origins of neutrino mass and dark matter (DM), keeping other avenues open for physics beyond the Standard Model (BSM). There have been several conclusive evidences in the last two decades which validate the existence of non-zero neutrino masses and large leptonic mixing \cite{PDG, kamland08, T2K, chooz, daya, reno, minos}. The present status of different neutrino parameters can be found in the latest global fit analysis \cite{schwetz16}. The SM can not address this observed phenomena simply because the neutrinos remain massless in the model. Due to the absence of the right handed neutrino, the Higgs field can not have any Dirac Yukawa coupling with the neutrinos. If the right handed neutrinos are included by hand, one needs the Yukawa couplings to be heavily fine tuned to around $10^{-12}$ in order to generate sub-eV neutrino masses from the same Higgs field of the SM. One can generate a tiny Majorana mass for the neutrinos from the same Higgs field of the SM at non-renormalisable level through the dimension five Weinberg operator \cite{weinberg}. The realisation of this dimension five operator within renormalisable theories are also available in the literature, popularly known as the seesaw mechanism \cite{ti}. Even if the tiny neutrino masses are generated dynamically within such seesaw frameworks, understanding the origin of the large leptonic mixing is another puzzle. Since the quark sector mixing is observed to be small, it also indicates that there may be some new dynamics operating in the leptonic sector that generates the large mixing. As can be seen from the global fit data, out of the three leptonic mixing angles, the solar and atmospheric angles are reasonably large while the reactor mixing angle is relatively small. In fact, before the discovery of non-zero reactor mixing angle $\theta_{13}$ in 2012, the neutrino data were consistent with a class of neutrino mass matrices obeying $\mu-\tau$ symmetry \footnote{For a recent review, please see \cite{xing2015}.}. This class of models predicts $\theta_{13} = 0, \theta_{23} = \frac{\pi}{4}$ whereas the value of $\theta_{12}$ depends upon the particular model. Out of different $\mu-\tau$ symmetric neutrino mass models, the Tri-Bimaximal (TBM) mixing \cite{Harrison} received lots of attention within several neutrino mass models. The TBM mixing predicts $\theta_{12}=35.3^o$. Such a mixing can be easily accommodated within popular discrete flavour symmetry models \cite{discreteRev}. Since the measured value of $\theta_{13}$ is small, such $\mu-\tau$ symmetric models can still be considered to be valid at leading order, while the small but non-zero $\theta_{13}$ can be generated by perturbations to either the charged lepton or the neutrino sector, as studied in several works in the literature including \cite{nzt13, nzt13A4, nzt13GA,db-t2, dbijmpa, dbmkdsp, dbrk}.

On the other hand, the SM also fails to provide a particle DM candidate that can satisfy all the criteria of a good DM candidate \cite{marco08}. Although there are enough evidences from astrophysics and cosmology suggesting the presence of DM, starting from the galaxy cluster observations by Fritz Zwicky \cite{zwicky} back in 1933, observations of galaxy rotation curves in 1970's \cite{rubin}, the more recent observation of the bullet cluster \cite{bullet} to the latest cosmology data provided by the Planck satellite \cite{par15}, the particle nature of DM is not yet known. This has motivated the particle physics community to study different possible BSM frameworks which can give rise to the correct DM phenomenology and can also be tested at several different experiments. Among them, the weakly interacting massive particle (WIMP) paradigm is the most popular BSM scenario as the correct DM relic abundance can be achieved for such a particle if it has interaction strength similar to weak interactions. This coincidence is also referred to as the \textit{WIMP Miracle}. In terms of density parameter and $h = \text{(Hubble Parameter)}/100$, the present dark matter abundance is conventionally reported as \cite{par15}
\begin{equation}
\Omega_{\text{DM}} h^2 = 0.1187 \pm 0.0017
\label{dm_relic}
\end{equation}
Using the measured value of Hubble parameter, this gives rise to approximately $26\%$ of the total energy density of the present Universe being made up of DM. The same Planck experiment also puts an upper bound on the lightest neutrino mass from the measurement of the sum of absolute neutrino masses $\sum_i \lvert m_i \rvert < 0.17$ eV \cite{par15}. Although the origin of neutrino mass as well as leptonic mixing may be unrelated to the fundamental origin of DM, it is highly motivating to look for a common framework that can explain both the phenomena. This not only keeps the BSM physics minimal, but also allows for its probe in a much wider range of experiments. We find two such frameworks very appealing: one where neutrino masses originate at one loop level with DM particles going in the loop \cite{ma06} and the other where the same discrete flavour symmetry responsible for generating large leptonic mixing also guarantees a stable DM candidate \cite{Valle10}. More detailed phenomenology of similar models can be found in several works including \cite{Peinado11,Ma08, Bouc12,Bouc11, am16, db16}. Another recent proposal to connect dark mater with non-zero $\theta_{13}$ can be found in \cite{arunansu}.

Motivated by this, here also we consider an inverse seesaw model \cite{inverse, Mohapatra86} based on $S_4$ discrete flavour symmetry that gives rise to TBM type neutrino mixing at leading order. Unlike canonical seesaw models, the inverse seesaw can be a low scale framework where the singlet heavy neutrinos can be at or below the TeV scale without any fine tuning of Yukawa couplings. This is possible due to softly broken global lepton number symmetry by the singlet mass term as we discuss later. The existence of sterile neutrinos around TeV scale with sizeable Yukawa couplings in these models makes these models testable at planned future particle colliders \cite{antush16}. Another motivation to study this particular model is the neutrino mass sum rules it predicts, which relates the three light neutrino masses \cite{Dorame12}. This predicts the lightest neutrino mass, once the experimental data of two mass squared differences are given as input and hence can be probed at experiments sensitive to the lightest neutrino mass say, neutrinoless double beta decay (NDBD) \footnote{For a review, please see \cite{NDBDrev}}. Since the model gives rise to TBM mixing, disallowed by latest neutrino data, we extend the model in order to generate non-zero $\theta_{13}$ in such a way that automatically takes DM into account. For this we make use of the scotogenic mechanism \cite{ma06} mentioned above where DM particles going in loop can generate tiny neutrino mass. We implement this idea in two different ways. First we add a one loop correction to the leading order light neutrino mass matrix from inverse seesaw and secondly we give a similar correction to the charged lepton mass matrix. In both the cases, the correct neutrino and DM phenomenology can be reproduced. However, the charged lepton correction is found to have advantage over the former due the fact that it does not disturb the mass sum rule prediction of the leading order model. Also, one requires less fine-tuning to generate correction to charged lepton masses due to which the lepton portal limit of inert scalar DM can be achieved, which can give different DM phenomenology compared to the well studied Higgs portal DM scenario, as we discuss later.

The work is organised as follows. In section \ref{sec:level2} we summarise the $S_4$ based inverse seesaw model at leading order along with its predictions. In section \ref{sec:level3} we explain the origin of non-zero reactor mixing angle and Dark Matter by extending the leading order model. In section \ref{sec:level4} we briefly discuss DM phenomenology of the model and then briefly comment upon neutrinoless double beta decay prediction in the context of the present model in section \ref{sec:level5}. We discuss our results in section \ref{sec:level6} and finally conclude in section \ref{sec:level7}.

\section{Inverse Seesaw Model with $S_4$ Symmetry}
\label{sec:level2}
In this section we briefly review the inverse seesaw model and its $S_4$ realisation. The inverse seesaw model is an extension of the SM by two different types of singlet neutral fermions $N_R, S_L$ three copies each. The Lagrangian is given by
\begin{equation}
-\mathcal{L} = Y \bar{L} h N_{R} + M \bar{S}_L N_{R} +\frac{1}{2}\mu  S_L S_L+\text{h.c.}
\end{equation}
Here $h$ is the SM Higgs doublet and $L$ is the lepton doublet. The presence of some additional symmetries is assumed which prevents the Majorana mass term of $N_R$. This Lagrangian gives rise to the following $9\times9$ mass matrix in the $(\nu_L, N_R, S_L)$ basis
            \begin{equation}\label{eq:2}
      M_{\nu}= \left(\begin{array}{ccc}
      0 & m^{T}_{D} & 0 \\
      m_{D}& 0 & M^{T}\\ 
      0 & M & \mu
      \end{array}\right)
      \end{equation}
where $m_D=Y\langle h^0 \rangle$ is the Dirac neutrino mass generated by the vacuum expectation value (vev) of the neutral component of the SM Higgs doublet. Block diagonalisation of the above mass matrix results in the effective light neutrino mass matrix as ,
       \begin{equation}\label{eq:3} 
        m_{\nu} = m_{D}^{T}(M^{T})^{-1} \mu M^{-1}m_{D}
       \end{equation}
Unlike canonical seesaw where the light neutrino mass is inversely proportional to the lepton number violating Majorana mass term of singlet neutrinos, here the light neutrino mass is directly proportional to the singlet mass term $\mu$. The heavy neutrino masses are proportional to $M$. Here, even if $M \sim 1$ TeV, correct neutrino masses can be generated for $m_D \sim 10$ GeV, say if $\mu \sim 1$ keV. Such small $\mu$ term is natural as $\mu \rightarrow 0$ helps in recovering the global lepton number symmetry $U(1)_L$ of the model. Thus, inverse seesaw is a natural TeV scale seesaw model where the heavy neutrinos can remain as light as a TeV and Dirac mass can be as large as the charged lepton masses and can still be consistent with sub-eV light neutrino masses.

In general, the inverse seesaw formula for light neutrino mass can generate a very general structure of neutrino mass matrix. Since the leptonic mixing is found to have some specific structure with large mixing angles, one can look for possible flavour symmetry origin of it. In this context, non Abelian discrete flavour symmetries have gained lots of attention in the last few decades. For reviews and related references, please see \cite{Alt10}. For the purpose of the present work, we are particularly interested in the inverse seesaw model proposed by \cite{Dorame12} where the non Abelian discrete flavour symmetry is $S_{4}$, the group of permutation of four objects, isomorphic to the symmetry group of a cube. The $S_{4}$ group has five irreducible representations, among which there are two singlets, one doublet and two triplets, the details of which are given in appendix \ref{appen1}. The field content of the $S_{4}$ based inverse seesaw model is shown in table \ref{tab2}. The additional discrete symmetry $Z_2\times Z_3$ as well as the global $U(1)_L$ symmetry is chosen in order to generate the desired inverse seesaw mass matrix along with TBM type leptonic mixing. The lepton doublet and charged lepton singlet of the SM, the singlet neutrinos $N_R, S$ of the inverse seesaw model transform as triplet $3_1$ of $S_4$. The SM Higgs doublet $h$ transform as singlet under $S_4$. The different flavon fields $\Phi$'s are chosen in order to get the desired mass matrices and mixing. The Yukawa Lagrangian for the particle content shown in table \ref{tab2} reads
\begin{equation} \label{eq:10}
        -\mathcal{L}^{I} =  y \bar{L}H N_{R} + y_{M}N_{R}S \Phi_{R} +  y_{M}^{\prime} N_{R}S \Phi_{R}^{\prime} +  y_{s} SS \Phi_{s} 
\end{equation}

        \begin{table}[htb]
       \begin{tabular}{|c|c|c|c|c|c|c|c|c|c|c|c|c|c|c|}
       \hline & $ \bar{L} $  & $ N_{R} $ & $ l_{R} $  & $H$  & $ S $  & $ \Phi_{R} $ & $\Phi_{R}^{\prime}$ & $\Phi_{s} $ & $\Phi_{l} $ & ${\Phi_{l}}^{\prime}
       $ & ${\Phi_{l}}^{\prime\prime} $  \\ 
       \hline $ SU(2)_{L} $ & 2  & 1 & 1 & 2 & 1  & 1 & 1  & 1 & 1 & 1 & 1 \\ 
       \hline $ S_{4} $ & $ 3_{1} $ &  $ 3_{1} $ & $ 3_{1} $ & $ 1_{1} $  & $ 3_{1} $  & $ 3_{1} $ & $ 1_{1} $  & $ 1_{1} $ & $3_{1}$ & $3_{2}$ & $1_{1}$  \\  
       \hline $ Z_{2} $ & + & + & + & + & - & - & -  & + & + & + & + \\
       \hline $ Z_{3} $ & $\omega^{2}$ & $\omega$ & 1 & 1 & 1 & $\omega^{2}$ & $\omega^{2}$ & 1 & $\omega$ & $\omega$ & $\omega$  \\
       \hline $U(1)_{L}$ & -1 & 1 & 1 & 0 & -1 & 0 & 0 & 2 & 0 & 0 & 0  \\
       \hline
       \end{tabular}
        \caption{Fields and their transformation properties under $ SU(2)_{L} $ gauge symmetry as well as the $ S_{4} \times Z_{2} \times Z_{3} \times U(1)_L$ symmetry} \label{tab2}
  \end{table} 
    The following flavon alignments are required to get a desired neutrino mass matrix and leptonic mixing.     
$$\langle \Phi_{R} \rangle = v_{R}(1,0,0), \; \langle \Phi_{R}^{\prime} \rangle = v_{R}^{\prime}, \;\langle \Phi_{s} \rangle = v_{s}, \; \langle H^0 \rangle =  v_{h}$$
    
     In order to implement this flavon alignment in the inverse seesaw mechanism 
     we note that $m_{D}$ is connected to $v_{h}$ and $M$ is determined by the vev $v_{R}$ and $v_{R}^{\prime}$. 
     In this way, the order
     of magnitude estimate of light neutrino mass from the equation (\ref{eq:3}) is $m_{\nu}\propto \frac{v_{h}^{2}}{(v_{R}+v_{R}^{\prime})^{2}}\mu$. Here $v_{h}$ is
     of the order of
     electroweak symmetry breaking (EWSB) scale, $v_{R}$ and $v_{R}^{\prime}$ can be taken of the order of TeV scale or more. Therefore, to get $m_{\nu}$ in sub-eV, $\mu$ which is coming 
     from the VEV of $\Phi_{S}$ should be of the order of keV. Such a small vev can be naturally achieved from the soft $U(1)_L$ symmetry breaking terms in the scalar potential. For example, a term $\mu_1 \Phi_s H^{\dagger} H$ will generate an induced vev of $\Phi_s$ given by $v_s = \frac{\mu_1 v^2_h}{M^2_{\Phi_s}}$. This can be adjusted to be keV by choosing a small enough $\mu_1$. By the same naturalness argument as before, such a small $\mu_1$ is natural. Also, since the $U(1)_L$ symmetry is explicitly broken (softly) by the scalar potential, there is no danger of generating massless Goldstone boson that can result after spontaneous breaking of global $U(1)_L$ symmetry.
   
  Decomposition of the various terms present in the equation (\ref{eq:10}) into singlets can be achieved using the $S_4$ tensor product rules given in appendix \ref{appen1} 
\begin{equation}
  y \bar{L}_{i}N_{jR}H = y (L_{1}N_{1R} + L_{2}N_{2R} + L_{3}N_{3R}) v_{h}
\end{equation}  
 \begin{align}
  y_M N_{iR}S_{j} \Phi_{R} &= y_{M} [(N_{2R}S_{3} + N_{3R}S_{2})\Phi_{1R} + (N_{1R}S_{3} + N_{3R}S_{1})\Phi_{2R} + (N_{1R}S_{2} + N_{2R}S_{1})\Phi_{3R}] \nonumber \\
  & =y_{M} [(N_{2R}S_{3} + N_{3R}S_{2})]v_{R} 
   \end{align}
      \begin{equation}
   y_{M}^{\prime} N_{iR}S_{j} \Phi_{R}^{\prime}  = y_{M}^{\prime} (S_{1}N_{1R} + S_{2}N_{2R} + S_{3}N_{3R})v_{R}^{\prime}
   \end{equation}
    \begin{equation}
  y_{s} SS \Phi_{s} = y_{s} (S_{1}S_{1} + S_{2}S_{2} + S_{3}S_{3})v_{s} 
  \end{equation}
   The chosen flavon alignments allow us to have different matrices involved in inverse seesaw formula as follows     
  \begin{equation} \label{eq:u} 
       m_{D}= y \left(\begin{array}{ccc}
       1 & 0 & 0 \\
       0 & 1 & 0\\ 
       0 & 0 & 1
       \end{array}\right)v_{h},\;  \mu  = y_{s}\left(\begin{array}{ccc}
        1 & 0 & 0 \\
       0 & 1 & 0\\ 
       0 & 0 & 1
        \end{array}\right)v_{s}, \;  M = \left(\begin{array}{ccc}
     y_{M}^{\prime} v_{R}^{\prime} & 0 & 0 \\
     0 & y_{M}^{\prime}v_{R}^{\prime} & y_{M}v_{R}\\ 
     0 & y_{M}v_{R} & y_{M}^{\prime}v_{R}^{\prime}
     \end{array}\right)
   \end{equation}
 The above three matrices lead to the following light neutrino mass matrix under ISS framework
    \begin{equation}\label{eq:b}
     m_{\nu} = U_{\nu}m_{\nu}^{o(diag)} U_{\nu}^{T}.
    \end{equation}
 Using \eqref{eq:u} in \eqref{eq:3}  the light neutrino mass matrix is found to be
   \begin{equation} \label{eq:q}
    m_{\nu}^o = \left(\begin{array}{ccc}
    \frac{1}{a^{2}} & 0 & 0 \\
    0 & \frac{a^{2}+b^{2}}{(b^2-a^2)^2} & -\frac{2ab}{(b^2-a^2)^2}\\ 
    0 & -\frac{2ab}{(b^2-a^2)^2} & \frac{a^{2}+b^{2}}{(b^2-a^2)^2}
    \end{array}\right)
    \end{equation}
    where, $a= y_{M}^{\prime}v_{R}^{\prime}/(\sqrt{y_{s}v_{s}} y v_{h})$ and $b= y_{M}v_{R}/(\sqrt{y_{s}v_{s}} y v_{h})$. The eigenvalues of this light neutrino mass matrix are 
$$ m_1 = \frac{1}{(a+b)^2}, \; m_2 = \frac{1}{(a-b)^2}, \; m_3 = \frac{1}{a^2} $$ 
which satisfy the neutrino mass sum rule
\begin{equation}
\frac{1}{\sqrt{m_1}} = \frac{2}{\sqrt{m_3}}-\frac{1}{\sqrt{m_2}}
\label{eq:sumrule}
\end{equation}
  Now the Lagrangian for the charged leptons can be written in terms of dimension five operators as \cite{Dorame12}
       \begin{equation}
        -\mathcal{L}^{l} =  \frac{y_{l}}{\Lambda}\bar{L}l_{R}H \Phi_{l} + \frac{{y_{l}}^{\prime}}{\Lambda}\bar{L}l_{R}H {\Phi_{l}}^{\prime} + 
        \frac{{y_{l}}^{\prime\prime}}{\Lambda}\bar{L}l_{R}H {\Phi_{l}}^{\prime\prime}
       \end{equation}
       The authors of \cite{Dorame12} considered additional messenger fields $\chi, \chi^c$ such that this effective Lagrangian for charged leptons can be obtained after integrating out these heavy messenger fields. The following flavon alignments allow us to have the desired mass matrix corresponding to the charged lepton sector
 $$\langle \Phi_{l} \rangle = v_{l}(1,1,1), \; \langle{\Phi_{l}}^{\prime}\rangle = {v_{l}}^{\prime}(1,1,1), \;\langle {\Phi_{l}}^{\prime\prime}\rangle = v_{l}^{\prime\prime}$$
 The charged lepton mass matrix is then given by
     \begin{equation}\label{eq:g}
      m^0_{l}= \left(\begin{array}{ccc}
   {y_{l}}^{\prime\prime}{v_{l}}^{\prime\prime} & y_{l}v_{l}-{y_{l}}^{\prime}{v_{l}}^{\prime} & 
   y_{l}v_{l}+{y_{l}}^{\prime}{v_{l}}^{\prime} \\
   y_{l}v_{l}+{y_{l}}^{\prime}{v_{l}}^{\prime} & {y_{l}}^{\prime\prime}{v_{l}}^{\prime\prime} &
   y_{l}v_{l}-{y_{l}}^{\prime}{v_{l}}^{\prime}\\ 
   y_{l}v_{l}-{y_{l}}^{\prime}{v_{l}}^{\prime} & y_{l}v_{l}+{y_{l}}^{\prime}{v_{l}}^{\prime} & {y_{l}}^{\prime\prime}{v_{l}}^{\prime\prime}
   \end{array}\right)\frac{v_h}{\Lambda}
     \end{equation}
As mentioned in \cite{Hirsch09} the charge lepton mass matrix $m_{l}$ is diagonalised on the left by the magic matrix $U_{\omega}$ given by 
  \begin{equation}\label{eq:x}
  U_{\omega} = 1/\sqrt{3} \left(\begin{array}{ccc}
   1 & 1 & 1 \\
   1 & \omega & \omega^{2}\\ 
   1 & \omega^{2} & \omega
   \end{array}\right),
    \end{equation}
    (with $\omega = \exp{2i\pi/3}$).
  Now we know that the leptonic mixing matrix is given by
  \begin{equation*}
   U=U_{\text{TBM}} = U_{l}^{\dag}U_{\nu}
  \end{equation*}
  where $U_{l}$ corresponds to the identity matrix if the charged lepton mass matrix is diagonal. Since in our work, the charged lepton mass matrix
  is non-diagonal and is nothing but the magic matrix $U_{\omega}$ given by \eqref{eq:x}, the leptonic mixing matrix is
  \begin{equation*}
  U_{\text{TBM}} = U_{\omega}^{\dag} U_{\nu}
  \end{equation*}

The desired structures of the mass and mixing matrices written above have been made possible due to chosen flavour symmetries of the theory.  For example, as required by the structure of the inverse seesaw mass matrix given in \eqref{eq:2}, there should not be any mass term involving $\nu_L$ and $S$. However, the coupling between $\nu_L$ and $S$ is not forbidden by the SM gauge symmetry as well as $S_4$ flavour symmetry.  In this regard, the additional $Z_2 \times Z_3$ symmetry and the chosen charges of $\nu_L, S$ under it keep the unwanted coupling of $\nu_L$ and $S$ through the Higgs doublet $H$ away. Similarly, the $(22)$ term of the inverse seesaw mass matrix \eqref{eq:2} or the mass term involving $N_R, N_R$ should also be forbidden. However, the SM gauge symmetry as well as the $S_4$ flavour symmetry and $U(1)_L$ global symmetry can not prevent a term like $\Phi_s N_R N_R$ which will introduce a non-zero $(22)$ entry into the inverse seesaw mass matrix. Therefore, the additional $Z_2 \times Z_3$ symmetry and non-trivial charges of $N_R$ under this has to be chosen to keep such a term away from the Lagrangian. As mentioned above, the approximate $U(1)_L$ global symmetry helps in generating small $(33)$ entry of the inverse seesaw mass matrix naturally, without any fine tuning of parameters. Thus, all the additional symmetries $Z_2 \times Z_3 \times U(1)_L$ play a crucial role in generating the desired structure of the inverse seesaw mass matrix along with the desired leptonic mixing.
\section{Origin of non-zero $\theta_{13}$ and Dark Matter}
\label{sec:level3}
Since $\theta_{13}=0$ has already been ruled out by several neutrino experiments, one has to go beyond the TBM framework discussed in the previous work. This can simply be done in two different ways: giving corrections to the neutrino mass matrix or the charged lepton mass matrix. Both of these corrections will change the leptonic mixing matrix in a way to generate non-zero $\theta_{13}$. 
\subsection{Correction to neutrino mass matrix}
The model discussed above can be extended by the particle content shown in table \ref{tab2b} charged under an additional $Z^{\text{Dark}}_2$ symmetry guaranteeing the stability of the dark matter candidate.   
\begin{table}[htb]
        \centering
       \begin{tabular}{|c|c|c|c|c|c|c|}   
        \hline & $ SU(2)_{L} $  & $ S_4 $ & $Z_2$ & $Z_3$ & $U(1)_L$ & $Z^{\text{Dark}}_2$ \\
         \hline $ \eta $ & 2  & 1 & 1 & 1 & 0 & -1 \\
         \hline $\psi_R$ & 1 & 3 & 1 & $\omega$ & 1 & -1 \\
        \hline $\Phi^{\prime}_{s}$ & 1 & 1 & 1 & $\omega$ & -2 & 1 \\ 
         \hline $\Phi_{\psi}$ & 1 & 3 & 1 & $\omega$ & -2 & 1 \\
\hline
\end{tabular}
\caption{Fields responsible for generating non-zero $\theta_{13}$ as well as dark matter with their respective transformations under the symmetry group of the model.} \label{tab2b}
\end{table}          
 \begin{figure}
\includegraphics[width=0.5\linewidth]{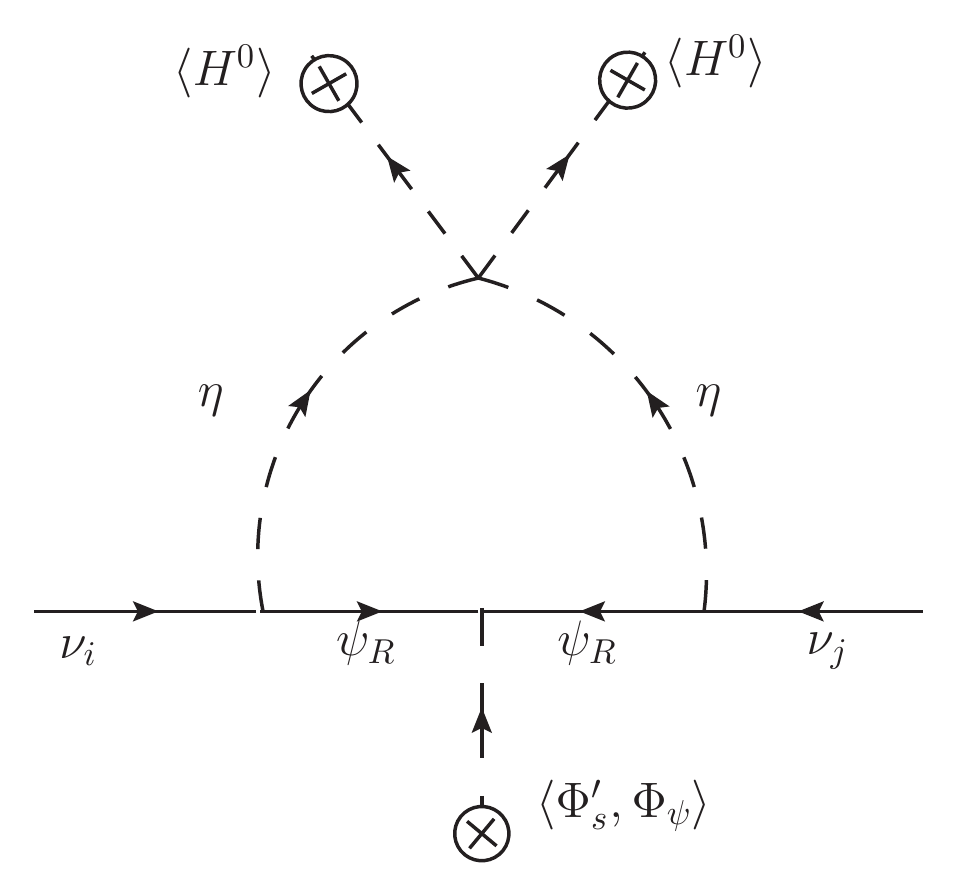}
\caption{Radiative generation of non-zero $\theta_{13}$ from the light neutrino sector}
\label{feyn1}
\end{figure} 
 This additional field content will introduce a few more terms in the Yukawa Lagrangian given as   
  \begin{equation} \label{eq:Ynew}
        \mathcal{L}^{I} \supset h \bar{L}\psi_{R} \eta +  y_{\psi} \psi_R \psi_R \Phi^{\prime}_{s} + y^{\prime} \psi_R \psi_R \Phi_{\psi}
       \end{equation}
The extra scalar doublet $\eta$ odd under the $Z^{\text{Dark}}_2$ symmetry introduces several other terms in the scalar potential. The most relevant terms are the interactions with the standard model Higgs $h$ which are relevant for neutrino mass and dark matter analysis. These relevant terms of the scalar potential can be written as
\begin{equation}
\begin{aligned}
V(H,\eta) \supset  \mu_1^2|H|^2 +\mu_2^2|\eta|^2+\frac{\lambda_1}{2}|H|^4+\frac{\lambda_2}{2}|\eta|^4+\lambda_3|H|^2|\eta|^2+\lambda_4|H^\dag \eta|^2 + \{\frac{\lambda_5}{2}(H^\dag \eta)^2 + \text{h.c.}\}
\end{aligned}
\label {c}
\end{equation}       
Using the expression from \cite{ma06} of one-loop neutrino mass 
\begin{equation}
(m_{\nu})_{ij} = \frac{h_{ik}h_{jk} M_{k}}{16 \pi^2} \left ( \frac{m^2_R}{m^2_R-M^2_k} \text{ln} \frac{m^2_R}{M^2_k}-\frac{m^2_I}{m^2_I-M^2_k} \text{ln} \frac{m^2_I}{M^2_k} \right)
\label{oneloop}
\end{equation}
Here $m^2_{R,I}$ are the masses of scalar and pseudoscalar part of $\eta^0$ and $M_k$ the mass of singlet fermion $\psi_{R}$ in the internal line. The index $i, j = 1,2,3$ runs over the three fermion generations as well as three copies of $\psi$. For $m^2_{R}+m^2_{I} \approx M^2_k$, the above expression can be simply written as
\begin{equation}
(m_{\nu})_{ij} \approx \frac{\lambda_5 v^2_h}{32 \pi^2}\frac{h_{ik} h_{jk} }{M_k} =  \frac{m^2_I-m^2_R}{32 \pi^2}\frac{h_{ik} h_{jk} }{M_k}
\end{equation}
where $m^2_I-m^2_R=\lambda_5 v^2_h$ is assumed ignoring the quartic terms of $\eta$ with other flavon fields. This formula for light neutrino mass is written in a basis where the mass matrix of the intermediate fermion $\psi$ is diagonal which is true if only $\Phi^{\prime}_s$ contributes to its mass $M_k = y_{\psi} \langle \Phi^{\prime}_{s} \rangle$ due to the structure of $S_4$ tensor product $\psi_R \psi_R \Phi^{\prime}_s = (\psi_{R1} \psi_{R1} +\psi_{R2} \psi_{R2} +\psi_{R3} \psi_{R3}) \Phi^{\prime}_s $. However, due to the $S_4$ triplet assignment to the other scalar $\Phi_{\psi}$, the mass matrix of $\psi_R$ becomes non-diagonal of the form
  \begin{equation}
  M_{\psi} =  \left(\begin{array}{ccc}
   y_{\psi} v^{\prime}_s & y^{\prime}_{\psi} v_{\psi 3} & y^{\prime}_{\psi} v_{\psi 2} \\
   y^{\prime}_{\psi} v_{\psi 3} & y_{\psi} v^{\prime}_s & y^{\prime}_{\psi} v_{\psi 1}\\ 
   y^{\prime}_{\psi} v_{\psi 2} & y^{\prime}_{\psi} v_{\psi 1} & y_{\psi} v^{\prime}_s
   \end{array}\right),
   \label{mpsi1}
    \end{equation}
where $\langle \Phi_{\psi} \rangle = (v_{\psi 1},v_{\psi 2},v_{\psi 3})$ is the vacuum alignment of the flavon field $\Phi_{\psi}$. Also the $S_4$ product rules dictate the Yukawa matrix $h_{ij}$ to be diagonal in flavour space. Therefore, the new contribution to the light neutrino mass matrix will assume a structure similar to $M_{\psi}$. We can parametrise this correction, in general as 
\begin{equation}\label{eq:qcorr}
\delta m_{\nu} = \left(\begin{array}{ccc}
x_{\nu} & y_{\nu} & z_{\nu} \\
y_{\nu} & x_{\nu} & w_{\nu} \\
z_{\nu} & w_{\nu} & x_{\nu} 
\end{array} \right)
\end{equation}
In this particular setup, the fermion $\psi_R$ carries lepton number, and since lepton number is only softly broken within an inverse seesaw framework, one expects the vev's of $\Phi^{\prime}_s, \Phi_{\psi}$ to be small say, of the order of keV in a TeV scale inverse seesaw model discussed above. Therefore, the dark matter in this model is a keV singlet fermion $\psi_R$. On the other hand, if $\psi_R$ does not carry a lepton number, then the scalar doublet $\eta$ carries a lepton number and the one-loop contribution can be generated with the particle content shown in table \ref{tab4}.
\begin{table}[ht!]
        \centering
       \begin{tabular}{|c|c|c|c|c|c|c|}   
        \hline & $ SU(2)_{L} $  & $ S_4 $ & $Z_2$ & $Z_3$ & $U(1)_L$ & $Z^{\text{Dark}}_2$ \\
         \hline $ \eta $ & 2  & 1 & 1 & 1 & 1 & -1 \\
         \hline $\psi_R$ & 1 & 3 & 1 & $\omega$ & 0 & -1 \\
         \hline $\Phi^{\prime}_{s}$ & 1 & 1 & 1 & $\omega$ & 0 & 1 \\ 
         \hline $\Phi_{\psi}$ & 1 & 3 & 1 & $\omega$ & 0 & 1 \\
         \hline $\Delta_{L}$ & 3 & 1 & 1 & 1 & 0 & 1 \\
\hline
\end{tabular}
\caption{Fields responsible for generating non-zero $\theta_{13}$ as well as dark matter with their respective transformations
under the symmetry group of the model.} \label{tab4}
\end{table}  
The Yukawa Lagrangian corresponding to this new field content is
  \begin{equation} \label{eq:Ynew2}
        \mathcal{L}^{I} \supset h \bar{L}\psi_{R} \eta +  y_{\psi} \psi_R \psi_R \Phi^{\prime}_{s} + y^{\prime} \psi_R \psi_R \Phi_{\psi}
       \end{equation}
These relevant terms of the scalar potential can be written as
\begin{equation}
\begin{aligned}
V(H,\eta, \Delta_L) & \supset  \mu_1^2|H|^2 +\mu_2^2|\eta|^2+\frac{\lambda_1}{2}|H|^4+\frac{\lambda_2}{2}|\eta|^4+\lambda_3|H|^2|\eta|^2+\lambda_4|H^\dag \eta|^2 \\
&+ \{\frac{\lambda_5}{2} \eta^2 \Delta_L \Phi_s + \text{h.c.}\},
\end{aligned}
\label {c1}
\end{equation}     
In this case, the fermion $\psi_R$ can acquire a diagonal mass term due to the coupling with $\Phi^{\prime}_{s}$ flavon and also acquire non diagonal mass terms from the flavon field $\Phi_{\psi}$. The combined mass matrix for $\psi_R$ therefore, has a similar structure to the one shown in equation \eqref{mpsi1}. Since neither $\psi_R$ nor $\Phi_{\psi}$ carries any lepton
number, their mass and vev respectively are not constrained to be small from naturalness argument. Also, the triplet scalar
$\Delta_L$ does not couple to the leptons at tree level as it does not carry any lepton number. The corresponding neutrino mass 
diagram at one loop is shown in figure \ref{feyn2}. This is equivalent to a radiative type II seesaw mechanism. In this case, the 
scalar doublet $\eta$ can be naturally lighter than $\psi_R$ and hence can be a dark matter candidate. We discuss this dark matter
candidate in details later, specially with reference to its interactions with the light neutrinos, responsible for generating 
non-zero $\theta_{13}$. In both these cases, the correction to the light neutrino mass matrix can be parametrised as \eqref{eq:qcorr}. One can then write down the complete light neutrino mass matrix as 
\begin{equation}
m_{\nu}=m^0_{\nu}+\delta m_{\nu} = U_{\text{PMNS}}m^{\text{diag}}_{\nu} U^T_{\text{PMNS}}
\label{nucorr1}
\end{equation}
where the Pontecorvo-Maki-Nakagawa-Sakata (PMNS) leptonic mixing matrix can be parametrized as
\begin{equation}
U_{\text{PMNS}}=\left(\begin{array}{ccc}
c_{12}c_{13}& s_{12}c_{13}& s_{13}e^{-i\delta}\\
-s_{12}c_{23}-c_{12}s_{23}s_{13}e^{i\delta}& c_{12}c_{23}-s_{12}s_{23}s_{13}e^{i\delta} & s_{23}c_{13} \\
s_{12}s_{23}-c_{12}c_{23}s_{13}e^{i\delta} & -c_{12}s_{23}-s_{12}c_{23}s_{13}e^{i\delta}& c_{23}c_{13}
\end{array}\right) U_{\text{Maj}}
\label{matrixPMNS}
\end{equation}
where $c_{ij} = \cos{\theta_{ij}}, \; s_{ij} = \sin{\theta_{ij}}$ and $\delta$ is the leptonic Dirac CP phase. The diagonal matrix $U_{\text{Maj}}=\text{diag}(1, e^{i\alpha}, e^{i(\beta+\delta)})$  contains the Majorana CP phases $\alpha, \beta$ which remain undetermined at neutrino oscillation experiments. For normal hierarchy, the diagonal mass matrix of the light neutrinos can be written  as $m^{\text{diag}}_{\nu} 
= \text{diag}(m_1, \sqrt{m^2_1+\Delta m_{21}^2}, \sqrt{m_1^2+\Delta m_{31}^2})$ whereas for inverted hierarchy 
 it can be written as $m^{\text{diag}}_{\nu} = \text{diag}(\sqrt{m_3^2+\Delta m_{23}^2-\Delta m_{21}^2}, 
\sqrt{m_3^2+\Delta m_{23}^2}, m_3)$. Using the $3\sigma$ values of neutrino parameters, we can find the model parameters in $m^0_{\nu}+\delta m_{\nu}$ which can give rise to the correct neutrino phenomenology.
\begin{figure}
\includegraphics[width=0.5\linewidth]{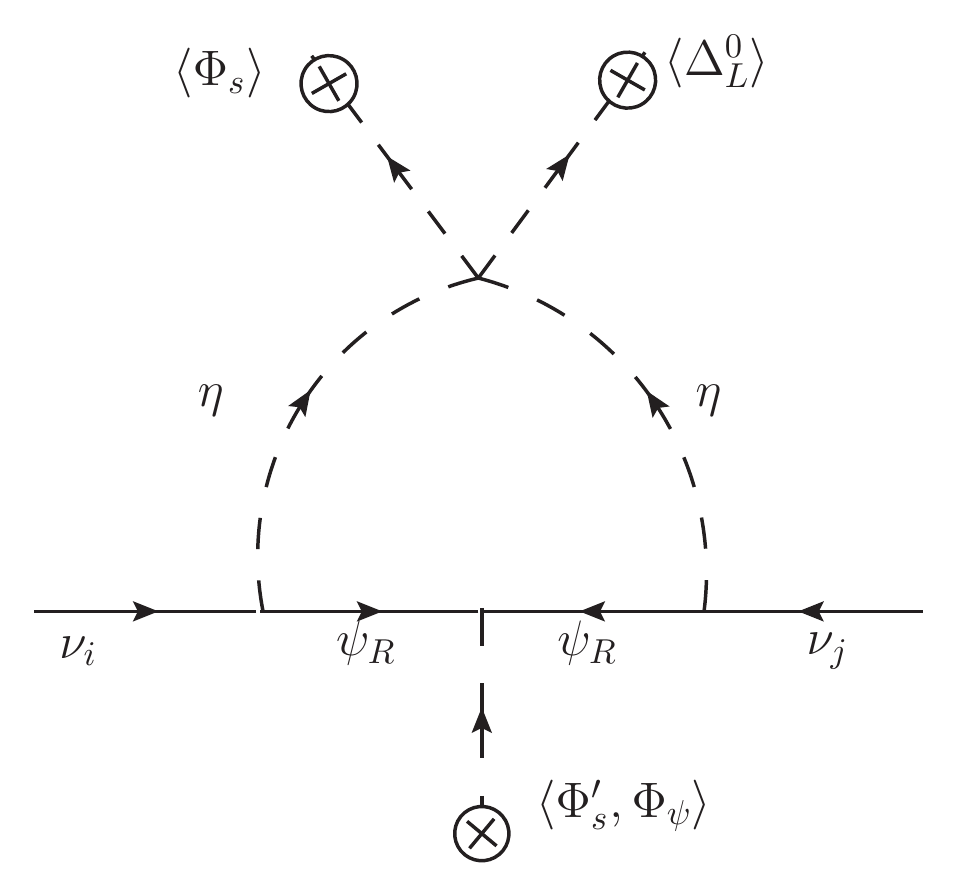}
\caption{Radiative generation of non-zero $\theta_{13}$ from the light neutrino sector}
\label{feyn2}
\end{figure} 
 \begin{figure}
\includegraphics[width=0.5\linewidth]{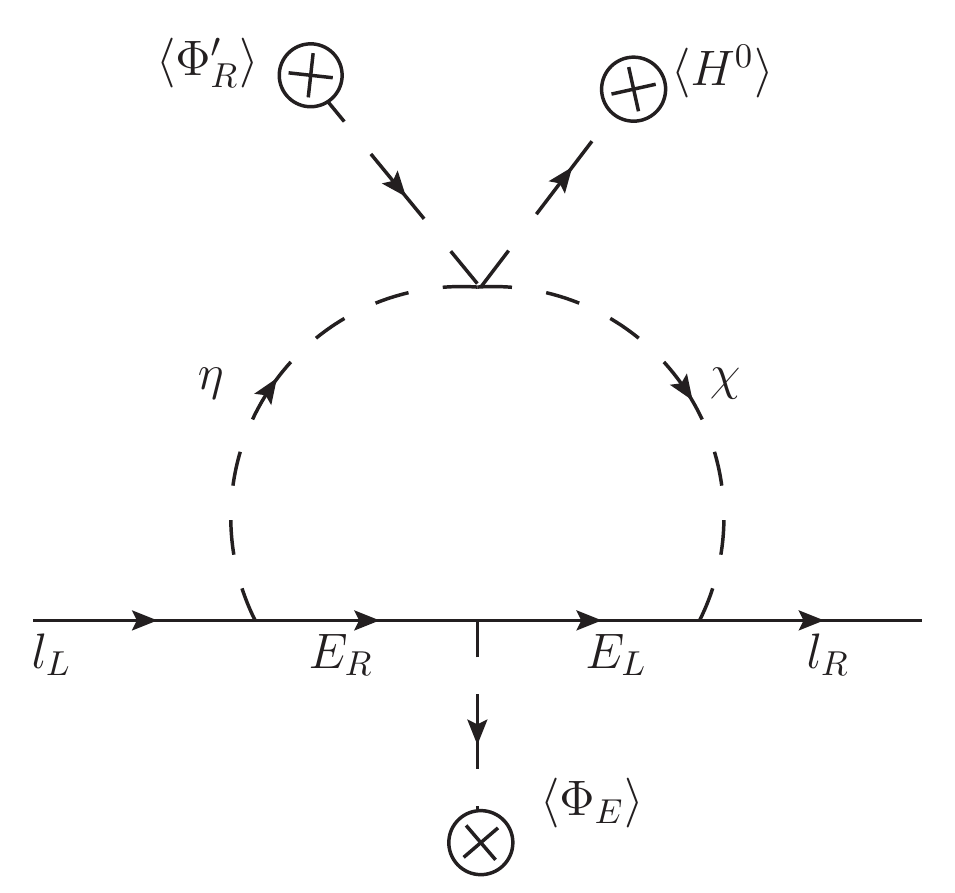}
\caption{Radiative generation of non-zero $\theta_{13}$ from charged lepton sector}
\label{feyn3}
\end{figure} 
\subsection{Correction to charged lepton mass matrix}
Similar to the above, one can also give a radiative correction to the charged lepton mass matrix, by considering the presence of vector like charged fermions instead of neutral ones. The relevant particle content is shown in table \ref{tab4b}. The Yukawa Lagrangian corresponding to this new field content is
  \begin{equation} \label{eq:Ynew3}
        \mathcal{L}^{I} \supset h \bar{L}E_{R} \eta^{\dagger} +h^{\prime} \bar{l}_RE_{L} \chi+ M_E \bar{E}_L E_R + y_E \Phi_E \bar{E}_L E_R 
       \end{equation}
These relevant terms of the scalar potential can be written as
\begin{equation}
\begin{aligned}
V & \supset  \mu_1^2|H|^2 +\mu_2^2|\eta|^2+\frac{\lambda_1}{2}|H|^4+\frac{\lambda_2}{2}|\eta|^4+\lambda_3|H|^2|\eta|^2+\lambda_4|H^\dag \eta|^2 \\
&+\{\frac{\lambda_5}{2}(H^\dag \eta)^2 + \text{h.c.}\} + \lambda_6 H^{\dagger} \eta \chi^{\dagger} \Phi^{\prime}_R
\end{aligned}
\label {c2}
\end{equation}     
The corresponding Feynman diagram for one-loop charged lepton mass is shown in figure \ref{feyn3}. One can write down the one-loop expression similar to the one written for one-loop neutrino masses. Here also, the mass matrix of vector like charged leptons acquire a similar structure as shown for neutral fermion $\psi_R$ in \eqref{mpsi1}. Also the Yukawa matrix related to the coupling of $\bar{l}_L E_R \eta$ or $\bar{l}_R E_L \chi$ is restricted to be diagonal due to $S_4$ product rules. Therefore, one can parametrise the correction to the charged lepton mass matrix as
\begin{equation}\label{eq:p}
\delta m_{l} = \left(\begin{array}{ccc}
a_l & b_l & c_l \\
b^s_l & a_l & d_l \\
c^s_l & d^s_l & a_l 
\end{array} \right)
\end{equation}
\begin{table}[htb]
        \centering
       \begin{tabular}{|c|c|c|c|c|c|c|}   
        \hline & $ SU(2)_{L} $  & $ S_4 $ & $Z_2$ & $Z_3$ & $U(1)_L$ & $Z^{\text{Dark}}_2$ \\
         \hline $ \eta $ & 2  & 1 & 1 & 1 & 0 & -1 \\
          \hline $ \chi $ & 1  & 1 & 1 & $\omega^2$ & 0 & -1 \\
         \hline $E_{L,R}$ & 1 & 3 & 1 & $\omega$ & 1 & -1 \\
         \hline $\Phi_{E}$ & 1 & 3 & 1 & 1 & 0 & 0 \\
         \hline
\end{tabular}
\caption{Fields responsible for generating non-zero $\theta_{13}$ as well as dark matter with their respective
transformations under the symmetry group of the model.} \label{tab4b}
\end{table}          
Adding this correction to the leading order charged lepton mass matrix given in equation \eqref{eq:g}
should give rise to a different diagonalising matrix $U_l$ of charged leptons. The structure of this matrix will
depend upon the parameters $a_l, b_l, c_l, d_l$ which can be constrained from the requirement of producing the correct leptonic mixing
matrix after multiplying with $U_{\nu}$, the diagonalising matrix of light neutrino mass matrix. From the tree level model one can find $U_{\nu} = U_{\omega} U_{\text{TBM}}$. Now, the total charged lepton mass matrix is 
\begin{equation}
m_l = m^0_l + \delta m_l = U_L m^{\text{diag}}_l U^{\dagger}_R
\label{CLeq}
\end{equation}
where $U_{L,R}$ are unitary matrices that can diagonalise the complex charged lepton mass matrix. Here $m^{\text{diag}}_l$ is the known diagonal charged lepton mass matrix. The unitary matrix $U_L$ goes into the observed leptonic mixing matrix and hence can be calculated as $U_L = U_{\nu} U^{\dagger}_{\text{PMNS}}$ which can be written in terms of known $U_{\nu}$ from the leading order model and the known PMNS mixing matrix. We parameterise the another unitary matrix $U_R$ in terms of three mixing angles and one phase and vary them randomly in $0-\pi/4$ for angles and $0-2\pi$ for phase. Thus, we can calculate the charged lepton mass matrix in terms of known parameters as well as randomly generated values of $U_R$. For each possible such charged lepton mass matrix, we can then solve the above equation \eqref{CLeq} and calculate the model parameters such that correct leptonic mixing can be achieved. In this model, the dark matter candidate
can either be a scalar doublet $\eta$ or a scalar singlet $\chi$. We discuss their dark matter phenomenology below specially with
reference to their interactions with the charged leptons.

 \section{Dark Matter} 
 \label{sec:level4}
In the very early epochs of the Universe, the abundance of a typical WIMP DM relic particle $(\eta)$ is usually taken to be the equilibrium abundance. When the
temperature of the radiation dominated Universe cools down below $T \sim m_{\eta}$, $\eta$ becomes non-relativistic and quickly after that it also decouples from
the thermal bath and its abundance freezes out. The final relic abundance of such a particle $\eta$ which was in thermal equilibrium at earlier epochs can be calculated by solving the Boltzmann equation    
    \begin{equation}
    \frac{dn_{\eta}}{dt} + 3Hn_{\eta} = - \langle \sigma v \rangle (n^{2}_{\eta}- (n^{\text{eqb}}_{\eta})^{2})
    \end{equation}
where $ n_{\eta} $ is the number density of the DM particle $ \eta $ and $ n^{\text{eqb}}_{\eta} $ is the equilibrium number density. Also, $H$ is the Hubble expansion rate of the Universe and $\langle \sigma v \rangle$ is the thermally averaged annihilation
    cross-section of the DM particle $ \eta $. It is clear from this equation that when $ \eta $ was in thermal equilibrium, the right hand side of it vanishes and the number density of DM decreases with time only due to the expansion of the Universe, as expected. The approximate analytical solution of the above Boltzmann equation gives \cite{Kolb:1990vq, kolbnturner}
\begin{equation}
\Omega_{\chi} h^2 \approx \frac{1.04 \times 10^9 x_F}{M_{Pl} \sqrt{g_*} (a+3b/x_F)}
\end{equation}
where $x_F = m_{\chi}/T_F$, $T_F$ is the freeze-out temperature, $g_*$ is the number of relativistic degrees of freedom at the time of freeze-out and $M_{Pl} \approx 10^{19}$ GeV is the Planck mass. Here, $x_F$ can be calculated from the iterative relation 
\begin{equation}
x_F = \ln \frac{0.038gM_{\text{Pl}}m_{\chi}<\sigma v>}{g_*^{1/2}x_F^{1/2}}
\label{xf}
\end{equation}
Typically, DM particles with electroweak scale mass and couplings freeze out at temperatures in the range $x_{F}\approx 20-30$. The expression for relic density also has a more simplified form given as \cite{Jungman:1995df}
\begin{equation}
\Omega_{\chi} h^2 \approx \frac{3 \times 10^{-27} \text{cm}^3 \text{s}^{-1}}{\langle \sigma v \rangle}
\label{eq:relic}
\end{equation}
In the model discussed in the previous section, there can be two different types of DM candidates, the lightest neutral particle under the $Z^{\text{Dark}}_2$ symmetry. In the model with corrections to neutrino sector, either the neutral fermion $\psi_R$ or the neutral component of the scalar doublet $\eta$ can be DM depending on their masses whereas in the latter model with corrections to the charged lepton sector, only the scalar DM is possible. To keep the discussion same for both these models, we briefly discuss scalar DM phenomenology in this work. The scalar DM relic abundance calculation has already been done in several works \cite{Barbieri:2006dq,Cirelli:2005uq,LopezHonorez:2006gr,honorez1,borahcline,DBAD14}. Typically, correct relic abundance can be satisfied for two regions of DM mass in such a model: one below the $W$ boson mass threshold and another around $550$ GeV or more. Here we focus mainly on the low mass regime where the dominant annihilation channel of DM is the one through Higgs portal interactions. Also, depending on the mass difference between different components of the scalar doublet $\eta$, coannihilations can also play a non-trivial role. In the limit where Higgs portal and coannihilation effects are sub-dominant, the DM can annihilate through the lepton portal interactions which are also relevant for correct neutrino phenomenology discussed above. We leave a detailed study of such lepton portal limit of scalar doublet DM to an upcoming work \cite{leptonPortal}. Here we briefly comment on the lepton portal interaction and its role in generating DM relic abundance using the approximate analytical formula mentioned above.
    
It is straightforward to see from the Lagrangian that the scalar DM can annihilate into leptons through a process mediated by heavy fermions $\psi$ or $E_{L,R}$. The corresponding annihilation cross-section is given by \cite{Bai14}
     \begin{equation}\label{eq:d}
     \sigma v = \frac{v^{2} h^{4} m ^{2}_{\eta}}{48\pi(m ^{2}_{\eta} + m ^{2}_{\psi})^{2}}
     \end{equation} 
 With $v\sim 0.3c$ is the typical relative velocity of the two DM particles at the freeze out temperature, $ \eta $ is the 
 relic particle (DM), $h$ is the Yukawa coupling, $ m_{\eta} $ the relic mass, $ m_{\psi} $ is the mass of the gauge singlet mediating the annihilation. We then vary the DM mass and the Yukawa coupling for different benchmark values of mediator masses and constrain the parameter space from the requirement of generating the correct DM relic abundance. It should be noted that, there are also constraints from DM direct detection experiments like LUX \cite{LUX16} which currently rules out DM-nucleon spin independent cross section above around $2.2 \times 10^{-46} \; \text{cm}^2$ for DM mass of around 50 GeV. However, the lepton portal interactions can not mediate DM-nucleon interactions and hence such bounds are weak in these cases. In fact, such null results at direct detection experiments will push lepton portal interactions of DM into a more favourable regime.
 \begin{figure}[ht!]
\centering
\includegraphics[width=0.3\textwidth]{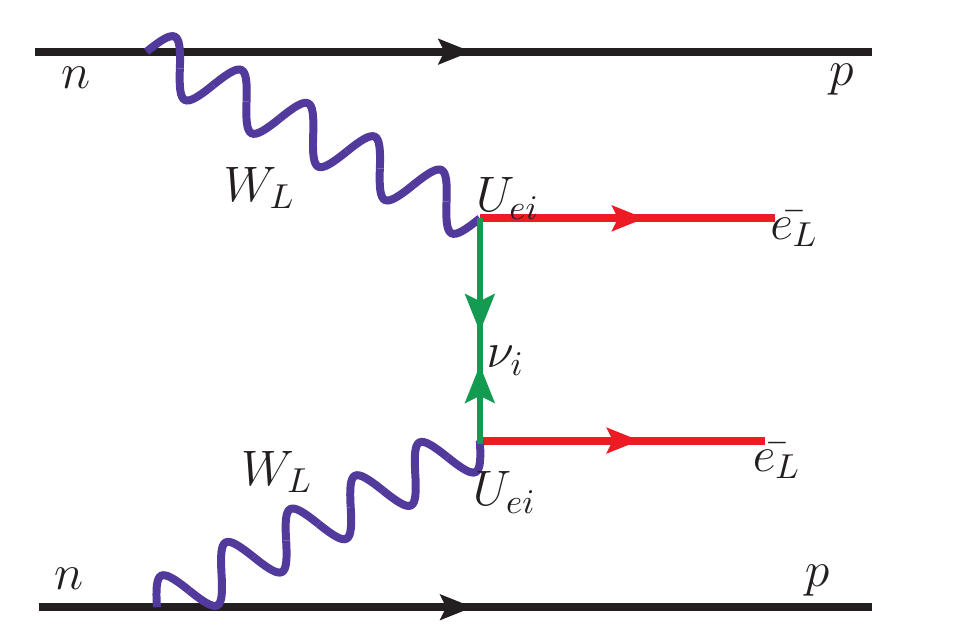}
\caption{Feynman diagram contributing to neutrinoless double beta decay due to light Majorana
 neutrino exchanges \cite{am16}.}
\label{ndbd}
\end{figure}

 \section{Neutrinoless Double Beta Decay}    
 \label{sec:level5}
The neutrinoless double beta decay (NDBD) is a lepton number violating process where a heavier nucleus decays into a lighter one and two electrons $(A, Z) \rightarrow (A, Z+2) + 2e^- $ without any antineutrinos in the final state. If the light neutrinos of SM are Majorana fermions, then they can contribute to NDBD through the interactions shown in the Feynman diagram of figure \ref{ndbd}. The amplitude of this light neutrino contribution is 
\begin{equation}
A_{\nu L L} \propto G^2_F \sum_i \frac{m_i U^2_{ei}}{p^2} 
\end{equation}
with $p$ being the average momentum exchange for the process. In the above expression, $m_i$ are the masses of light neutrinos for $i=1,2,3$ and $U$ is the PMNS leptonic mixing matrix mentioned earlier. The corresponding half-life of neutrinoless double beta decay can be written as 
\begin{align}
\frac{1}{T^{0\nu}_{1/2}} = G^{0\nu}_{01} \bigg ( \lvert \mathcal{M}^{0\nu}_\nu  (\eta^L_{\nu})\rvert^2 \bigg )
\end{align}
where $ \eta^L_{\nu} = \sum_i \frac{m_i U^2_{ei}}{m_e}$ with $m_e$ being the mass of electron. Also, $\mathcal{M}^{0\nu}_\nu$ is the nuclear matrix element. The recent bound from the KamLAND-Zen experiment constrains $0\nu \beta \beta$ half-life \cite{kamland2}
$$ T^{0\nu}_{1/2} (\text{Xe}136) > 1.1 \times 10^{26} \; \text{yr} $$
which is equivalent to $\lvert M^{ee}_{\nu} \rvert <(0.06 - 0.16)$ eV at $90\%$ C.L. where $M^{ee}_{\nu}$ is the effective neutrino mass given by
\begin{equation}
 M^{ee}_{\nu} = U^{2}_{ei} m_{i}
\end{equation}
Here $U_{ei}$ are the elements of the first row of the PMNS mixing matrix. More explicitly, it is given by
\begin{equation}
 M^{ee}_{\nu} = m_1c^2_{12}c^2_{13}+m_2s^2_{12}c^2_{13}e^{2i\alpha}+m_3 s^2_{13}e^{2i\beta} 
 \end{equation} 
Thus, the NDBD half-life is sensitive to the Majorana phases and the lightest neutrino mass as well, which remain undetermined at neutrino oscillation experiments. In the present model, the light neutrino contribution is the only dominant contribution. We check the predictions of our model for NDBD effective mass for both the cases and compare with the experimental bounds.
   
\begin{center}
\begin{table}[htb]
\begin{tabular}{|c|c|c|}
\hline
Parameters & Normal Hierarchy (NH) & Inverted Hierarchy (IH) \\
\hline
$ \frac{\Delta m_{21}^2}{10^{-5} \text{eV}^2}$ & $7.03-8.09$ & $7.02-8.09 $  \\
$ \frac{|\Delta m_{3l}^2|}{10^{-3} \text{eV}^2}$ & $2.407-2.643$ & $2.399-2.635 $  \\
$ \sin^2\theta_{12} $ &  $0.271-0.345 $ & $0.271-0.345 $  \\
$ \sin^2\theta_{23} $ & $0.385-0.635$ &  $0.393-0.640 $  \\
$\sin^2\theta_{13} $ & $0.01934-0.02392$ & $0.01953-0.02408 $  \\
$ \delta $ & $0^{\circ}-360^{\circ}$ & $145^{\circ}-390^{\circ}$  \\
\hline
\end{tabular}
\caption{Global fit $3\sigma$ values of neutrino oscillation parameters \cite{schwetz16}. Here $\Delta m_{3l}^2 \equiv \Delta m_{31}^2$ for NH and $\Delta m_{3l}^2 \equiv \Delta m_{32}^2$ for IH.}
\label{tab:gfit}
\end{table}
\end{center}

\section{Results and Discussions} 
\label{sec:level6} 
\begin{figure*}
\begin{center}
\includegraphics[width=0.45\textwidth]{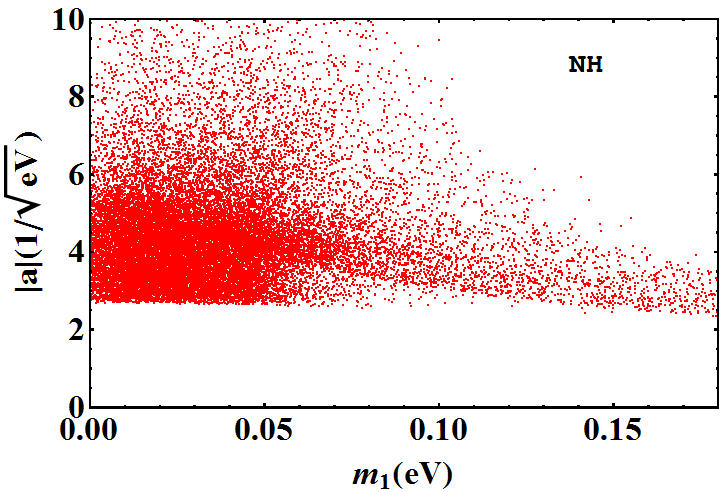}
\includegraphics[width=0.45\textwidth]{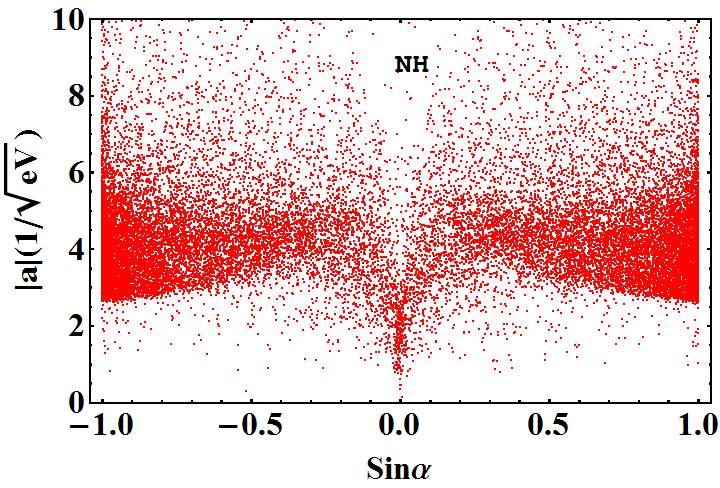} \\
\includegraphics[width=0.45\textwidth]{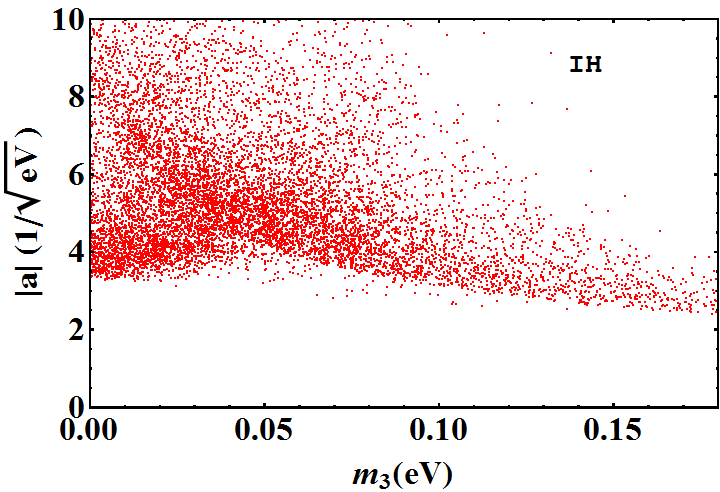}
\includegraphics[width=0.45\textwidth]{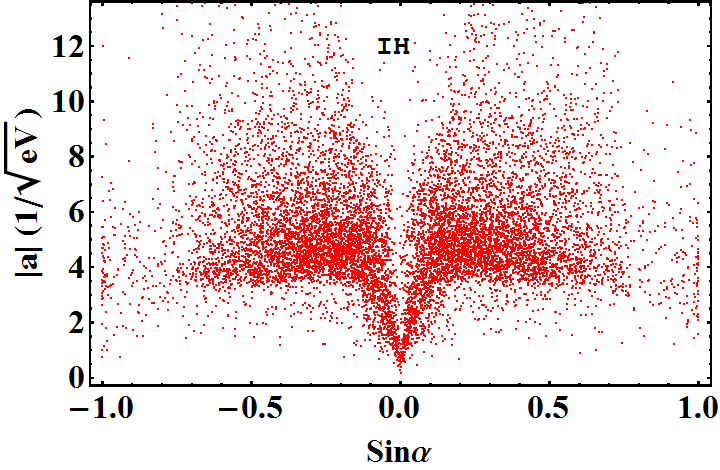}
\end{center}
\begin{center}
\caption{Model parameter as a function of the lightest neutrino mass and Majorana phase $\alpha$.}
\label{fig5}
\end{center}
\end{figure*}

\begin{figure*}
\begin{center}
\includegraphics[width=0.45\textwidth]{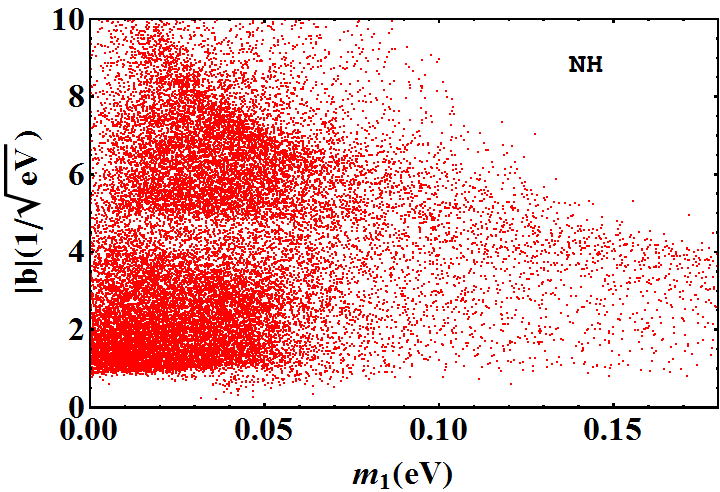}
\includegraphics[width=0.45\textwidth]{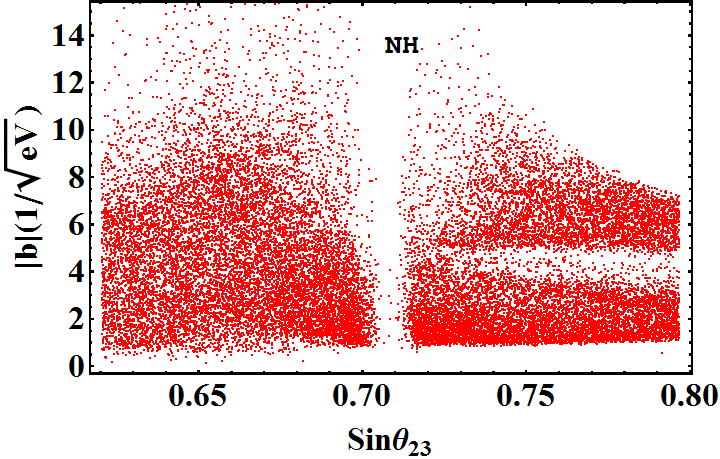} \\
\includegraphics[width=0.45\textwidth]{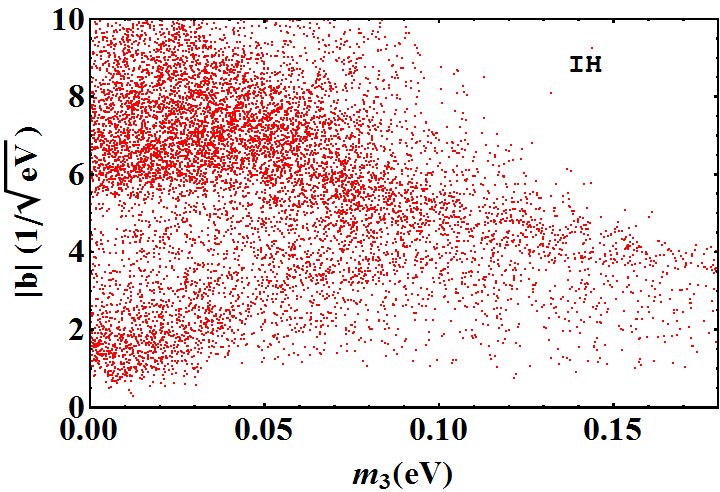}
\includegraphics[width=0.45\textwidth]{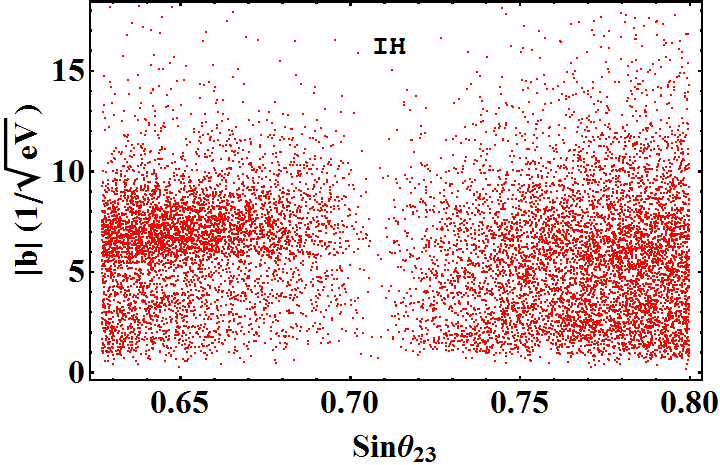}
\end{center}
\begin{center}
\caption{Model parameters as a function of the lightest neutrino mass and the atmospheric mixing angle $\theta_{23}$.}
\label{fig6}
\end{center}
\end{figure*}

\begin{figure*}
\begin{center}
\includegraphics[width=0.45\textwidth]{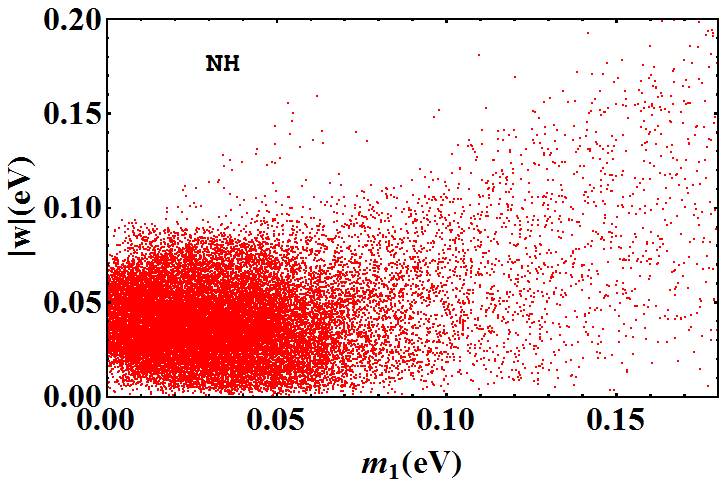}
\includegraphics[width=0.45\textwidth]{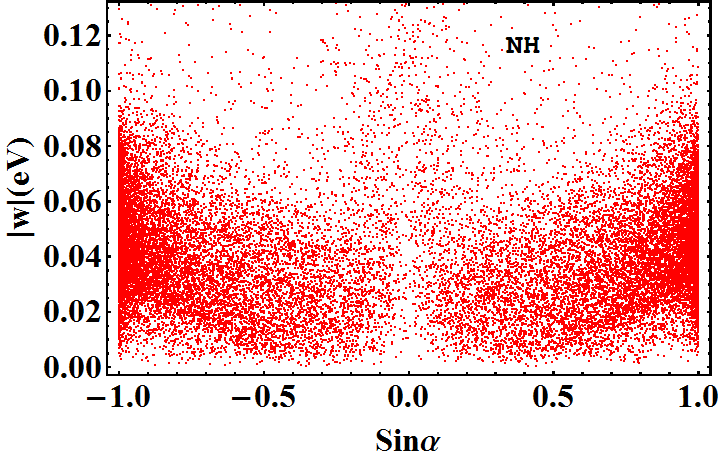} \\
\includegraphics[width=0.45\textwidth]{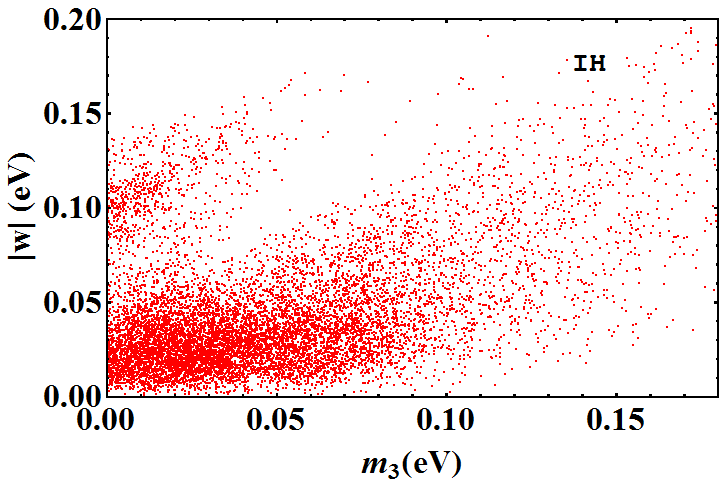}
\includegraphics[width=0.45\textwidth]{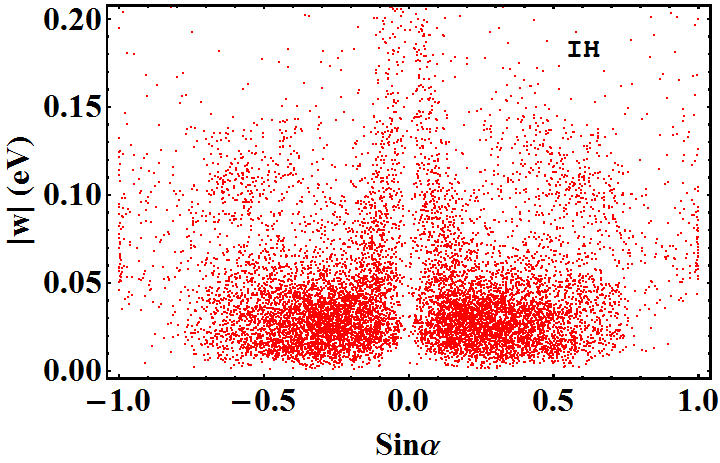}
\end{center}
\begin{center}
\caption{Corrections parameter(correction to neutrino mass matrix) as a function of lightest neutrino mass and Majorana phase $\alpha$.}
\label{fig7}
\end{center}
\end{figure*}

\begin{figure*}
\begin{center}
\includegraphics[width=0.45\textwidth]{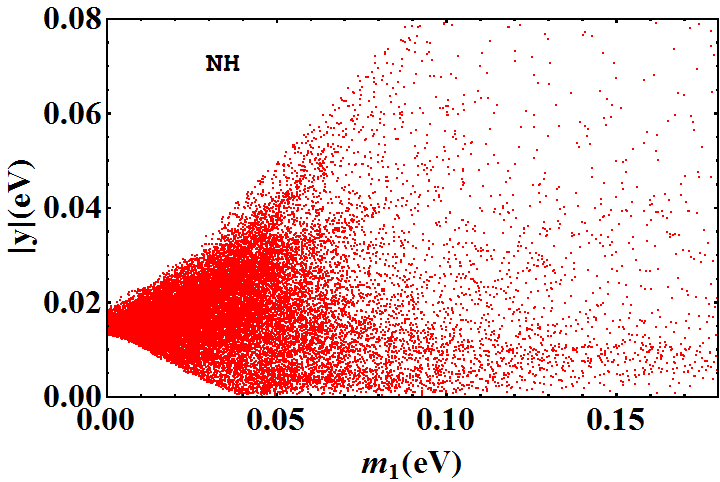}
\includegraphics[width=0.45\textwidth]{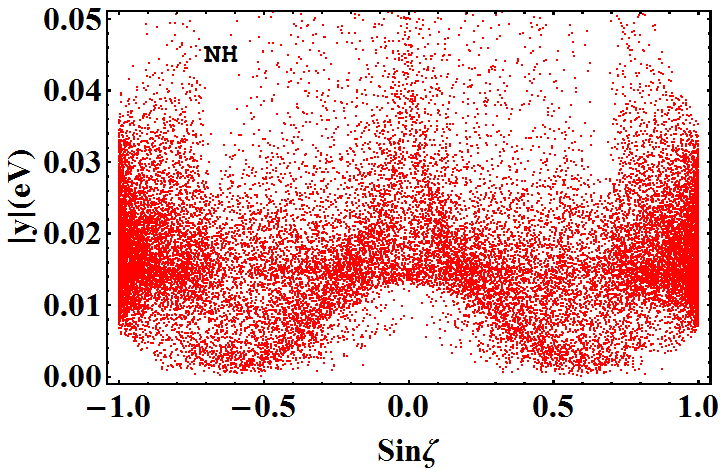} \\
\includegraphics[width=0.45\textwidth]{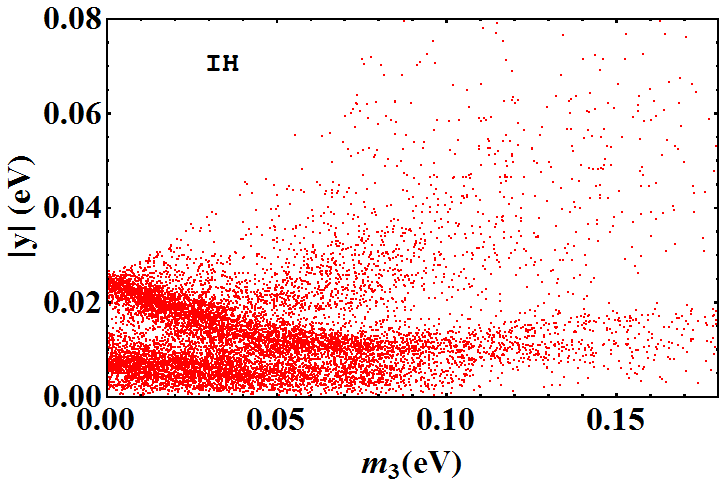}
\includegraphics[width=0.45\textwidth]{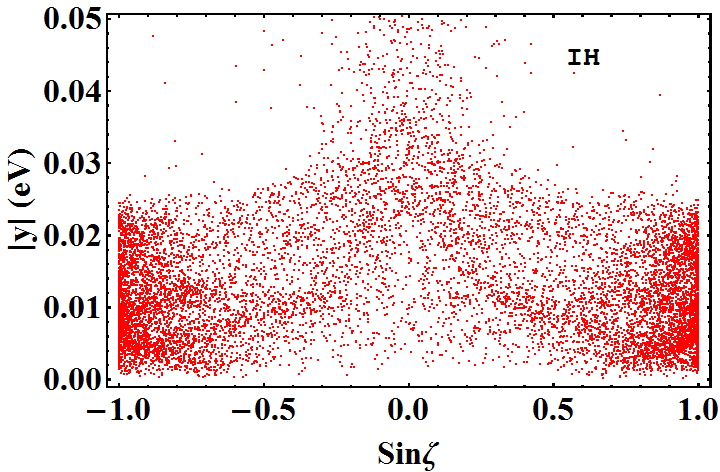}
\end{center}
\begin{center}
\caption{Corrections parameter(correction to neutrino mass matrix) as a function of lightest neutrino mass and Majorana phase $\zeta$.}
\label{fig8}
\end{center}
\end{figure*}

\begin{figure*}
\begin{center}
\includegraphics[width=0.45\textwidth]{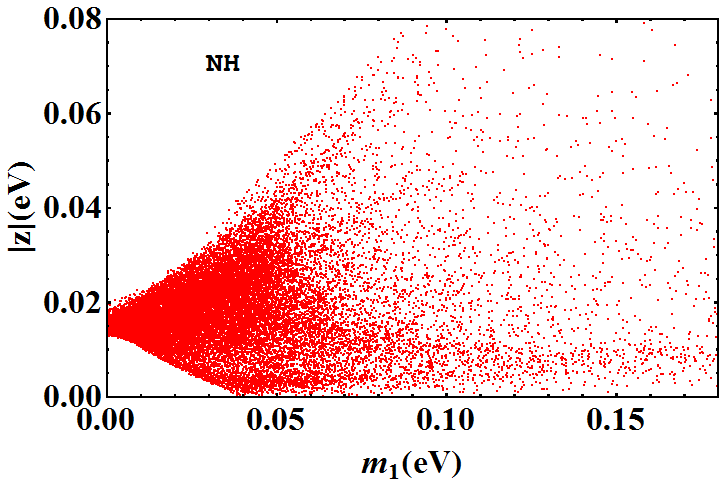}
\includegraphics[width=0.45\textwidth]{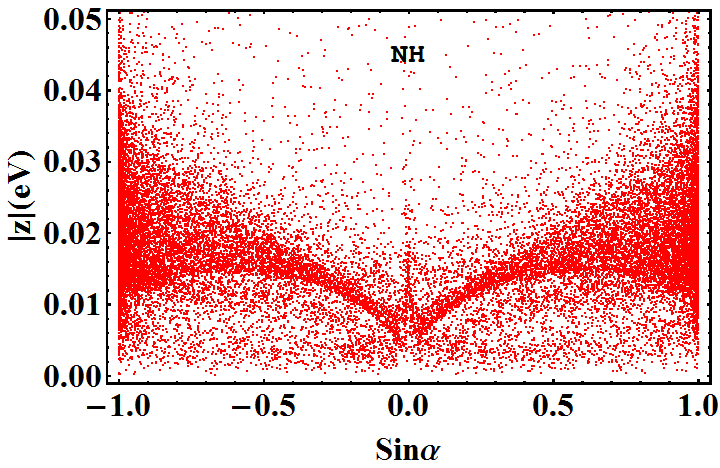} \\
\includegraphics[width=0.45\textwidth]{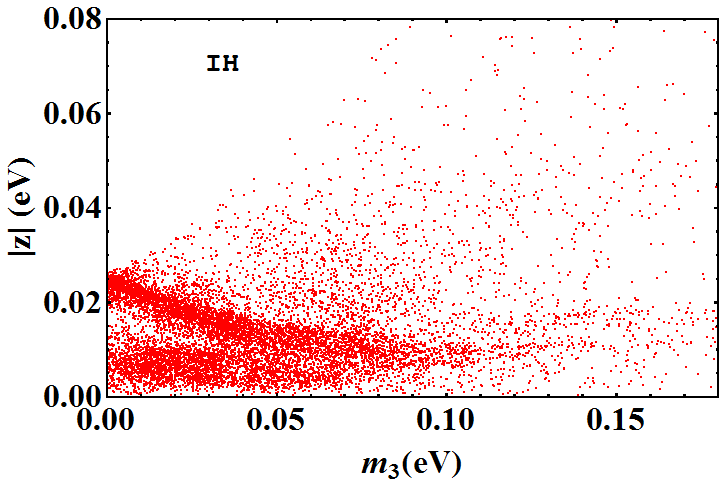}
\includegraphics[width=0.45\textwidth]{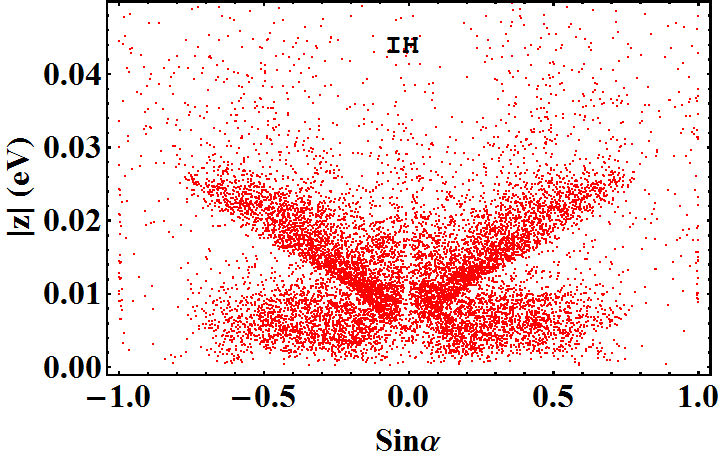}
\end{center}
\begin{center}
\caption{Corrections parameter(correction to neutrino mass matrix) as a function of lightest neutrino mass and Majorana phase $\alpha$.}
\label{fig9}
\end{center}
\end{figure*}

We first parametrize the light neutrino mass matrix in terms of the $3\sigma$ global fit data available \cite{schwetz16} which are summarised in table \ref{tab:gfit}. For the correction to the neutrino sector case, we then use \eqref{nucorr1} to relate the light neutrino mass matrix predicted by the model with the one parametrized by the global fit data. The leading order neutrino mass matrix given by \eqref{eq:q} contains two complex parameters $a, b$ whereas the correction to light neutrino mass is made up of four complex parameters $x, y, z, w$ as seen from \eqref{eq:qcorr}. The parametric form of light neutrino mass matrix is complex symmetric and hence contains six complex elements. Therefore, one can exactly solve the system of equations arising from \eqref{nucorr1} in order to evaluate the model parameters in terms of the known neutrino parameters. To be more precise, there are in fact five complex equations and one constraints arising from \eqref{nucorr1}. This is due to the fact that in the total neutrino mass matrix predicted by the model, we have the 22 and 33 entries equal. This in fact restricts the light neutrino parameters, as it gives rise to two real equations involving the light neutrino parameters. We first solve these system of equations and generate the light neutrino parameters which satisfy them. For the resulting light neutrino parameters, we solve the other five complex equations to evaluate the model parameters. Since we have six model parameters and only five equations now, we vary the parameter $x$ in the correction term \eqref{eq:qcorr} randomly in a range $10^{-6}-10^{-1}$ eV. Since there are nine neutrino parameters namely, three masses, three angles and three phases, one can in general, show the variation of model parameters in terms of all of these nine parameters which are being varied randomly in their allowed ranges. Here we show only a few of them for illustrative purposes. For example, we show the variation of some of the model parameters in terms of the light neutrino parameters in figure \ref{fig5}, \ref{fig6}, \ref{fig7}, \ref{fig8} and \ref{fig9}. This shows that the model parameters in the leading order and the correction mass matrices can not be arbitrary, but have to be within some specific ranges in order to be consistent with correct light neutrino data. From the figures \ref{fig5} and \ref{fig6} it is seen that the parameters of the leading order light neutrino mass matrix are in the range $a, b \approx 1-10 \; \text{eV}^{-1/2}$. We recall the expressions for $a, b$ in terms of the model parameters $a= y_{M}^{\prime}v_{R}^{\prime}/(\sqrt{y_{s}v_{s}} y v_{h})$ and $b= y_{M}v_{R}/(\sqrt{y_{s}v_{s}} y v_{h})$ mentioned earlier. Taking the lepton number violating term $\mu  =y_{s}v_{s} \approx 1 \; \text{keV}$, the vev of Higgs doublet at electroweak scale $v_h \approx 100 \; \text{GeV}$ and the vev of the other scalars $\Phi_R, \Phi^{\prime}_R$ around a TeV that is, $ v_R, v^{\prime}_R \approx 1 \; \text{TeV}$, our numerical results suggest that 
\begin{equation}
\frac{y_M}{y}=\frac{y^{\prime}_M}{y} \approx 10-1000
\end{equation}
in order to satisfy the correct neutrino data. This can be achieved by suitable tuning of the Dirac Yukawa $y$ relative to $y_M=y^{\prime}_M$. On the other hand, from the figures \ref{fig7}, \ref{fig8} and \ref{fig9}, it can be seen that the correction terms to the light neutrino mass matrix lie in the sub-eV regime. The one loop correction term shown in equation \eqref{oneloop} can be approximated for $m^2_{R}+m^2_{I} \approx M^2_k$, the above expression can be simply written as
\begin{equation}
(m_{\nu})_{ij} \approx \frac{\lambda_5 v^2_h}{32 \pi^2}\frac{h_{ik}h_{jk} }{M_k} =  \frac{m^2_I-m^2_R}{32 \pi^2}\frac{h_{ik}h_{jk} }{M_k}
\end{equation}
If the heavy neutrino mass $M_k$ is around a TeV, then for $m^2_I-m^2_R \approx 1 \; \text{GeV}$, one can generate sub eV scale corrections $\sim 0.01\; \text{eV}$ if the corresponding Yukawa couplings are fine tuned to $h \approx 10^{-3}$.


\begin{figure*}
\begin{center}
\includegraphics[width=0.45\textwidth]{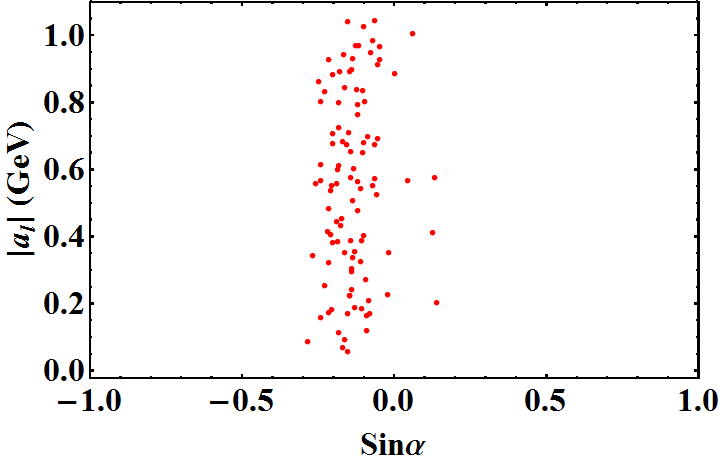}
\includegraphics[width=0.45\textwidth]{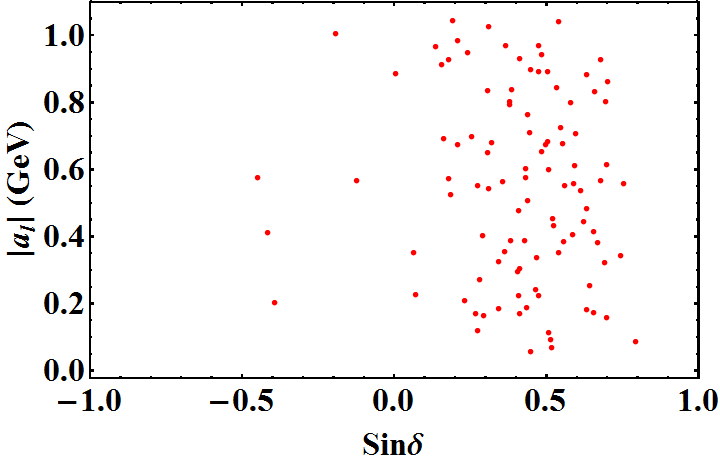} \\
\includegraphics[width=0.45\textwidth]{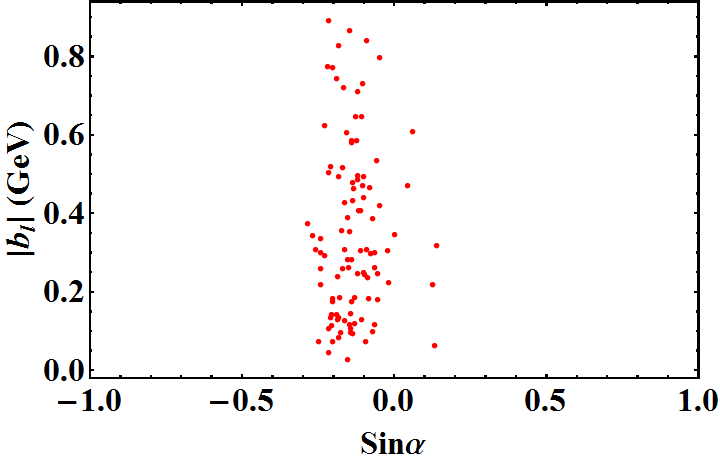}
\includegraphics[width=0.45\textwidth]{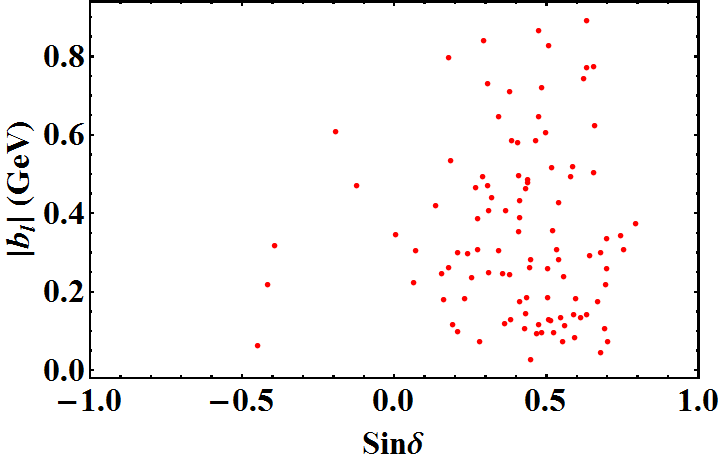}
\end{center}
\begin{center}
\caption{Correction parameters as a function of Majorana and Dirac phases while giving correction to the charged lepton mass matrix.}
\label{fig10}
\end{center}
\end{figure*}

\begin{figure*}
\begin{center}
\includegraphics[width=0.45\textwidth]{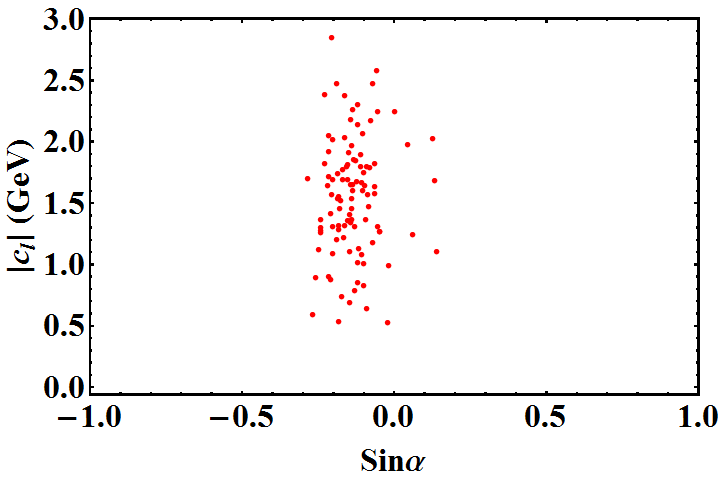}
\includegraphics[width=0.45\textwidth]{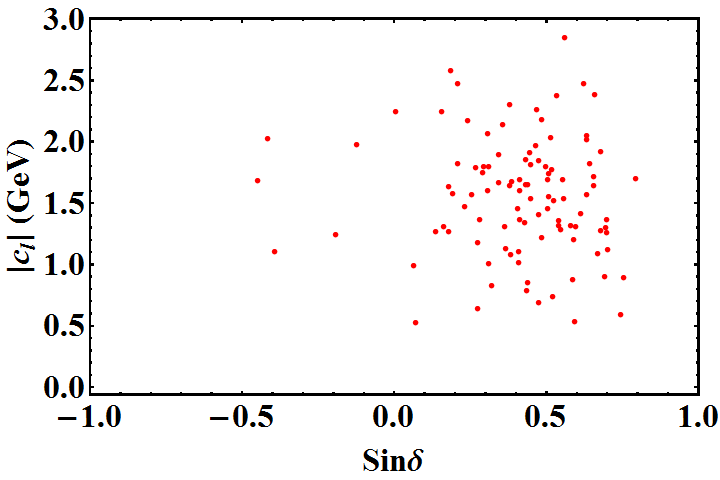} \\
\includegraphics[width=0.45\textwidth]{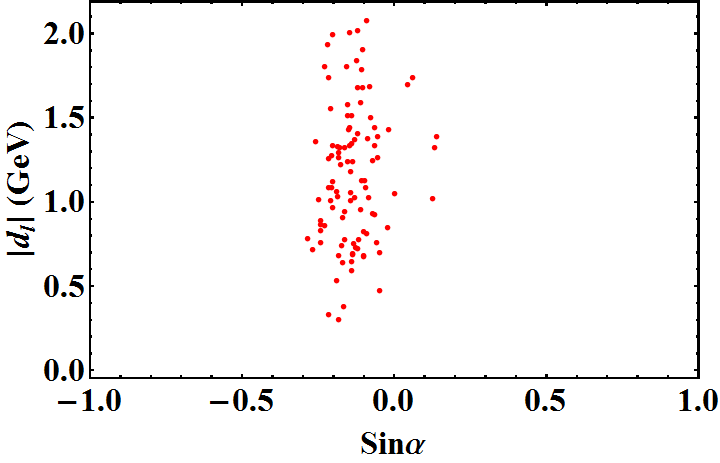}
\includegraphics[width=0.45\textwidth]{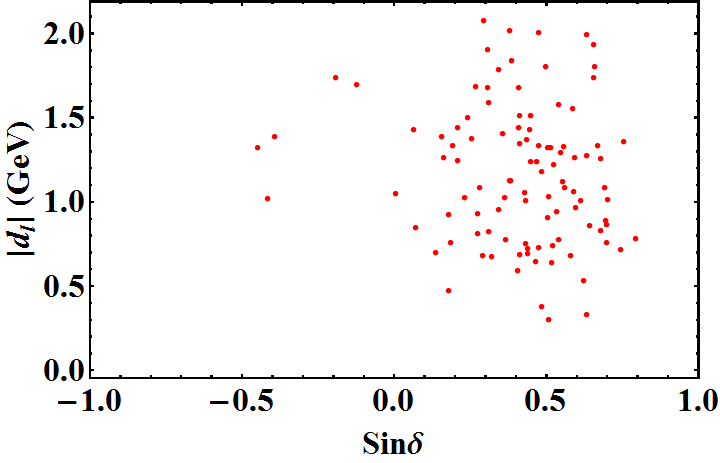}
\end{center}
\begin{center}
\caption{Correction parameters as a function of Majorana and Dirac phases while giving correction to the charged lepton mass matrix.}
\label{fig11}
\end{center}
\end{figure*}

In the model with corrections to the leading order charged lepton mass matrix, we first find out the diagonalising matrix of light neutrino mass matrix as $U_{\nu} = U_{\omega} U_{\text{TBM}}$ using the leading order results mentioned before. Since the light neutrino mass matrix remains the same after the charged lepton correction, $U_{\nu}$ also remains same. However the addition of correction will change the left diagonalising matrix of charged lepton mass matrix from the magic matrix $U_{\omega}$ to something else, denoted by $U_L = U_{\nu} U^{\dagger}_{\text{PMNS}}$. Now, using \eqref{CLeq}, one can relate the complete charged lepton mass matrix predicted by the model, with the parametrized one given by the right hand side of \eqref{CLeq}. The total charged lepton mass matrix can be written as 
\begin{equation}\label{CLcorr1}
m_l=m^0_l+\delta m_{l} = \left(\begin{array}{ccc}
x+a_l & y-z+b_l & y+z+c_l \\
y+z+b^s_l & x+a_l & y-z+d_l \\
y-z+c^s_l & y+z+d^s_l & x+a_l 
\end{array} \right)
\end{equation}
which contains ten complex parameters. Here $x,y,z$ correspond to ${y_{l}}^{\prime\prime}{v_{l}}^{\prime\prime}, y_{l}v_{l}, {y_{l}}^{\prime}{v_{l}}^{\prime} $ respectively in the leading order charged lepton mass matrix \eqref{eq:g}. Also there are two constraints in the parametrized charged lepton mass matrix due to fact that the $11$, $22$ and $33$ elements are equal. This severely constraints the mixing angles and phases. Since the angles contained in $U_L$ are related to the PMNS mixing angles, they can not be tuned arbitrarily. This forces some of the angles in $U_R$ to take very small values in order to satisfy these two constraints. The tiny values are required in order to compensate for the large hierarchy in charged lepton masses which enters the $11$, $22$ and $33$ elements of the mass matrix. We first solve these constraints numerically and then find the model parameters for those allowed values of mixing angles. We vary $x, y, z$ randomly in $10^{-6}-1.0$ GeV and evaluate other model parameters $a_l, b_l, c_l, d_l, b^s_l, c^s_l, d^s_l$ from the requirement of producing the correct leptonic mixing data. Unlike the earlier model with corrections to the neutrino mass matrix, here we get very few number of allowed points. For illustrative purposes we show the variation of $a_l, b_l, c_l, d_l$ with some light neutrino parameters in figure \ref{fig10} and \ref{fig11}. Since these one loop correction terms lie in the sub GeV regime, one can generate them without much fine tuning in the corresponding Yukawa couplings.

\begin{figure*}
\includegraphics[width=0.75\linewidth]{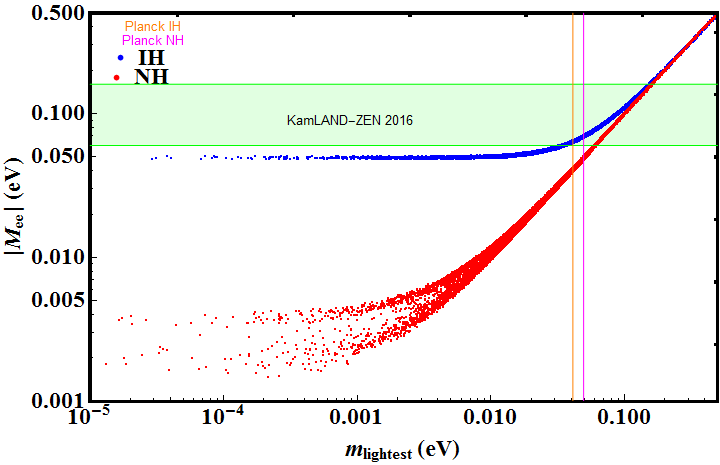}
\caption{Variation of effective neutrino mass with the lightest neutrino mass in the model with neutrino mass correction. The purple line indicates the PLANCK bound on the sum of absolute neutrino masses. The green band shows the KamLAND-ZEN upper bound \cite{kamland2} on the effective neutrino mass.}
\label{fig12}
\end{figure*} 

\begin{figure*}
\includegraphics[width=0.75\linewidth]{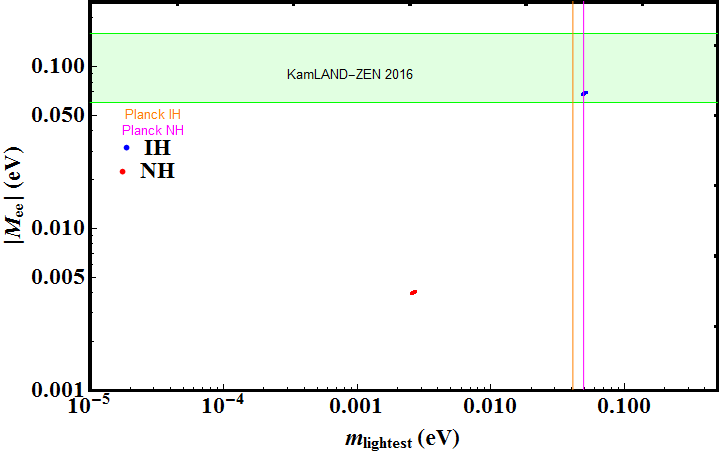}
\caption{Variation of effective neutrino mass with the lightest neutrino mass in the model with charged lepton correction. The purple line indicates the PLANCK bound on the sum of absolute neutrino masses. The green band shows the KamLAND-ZEN upper bound \cite{kamland2} on the effective neutrino mass.}
\label{fig13}
\end{figure*} 

\begin{figure*}
    \centering
    \includegraphics[width=0.45\textwidth]{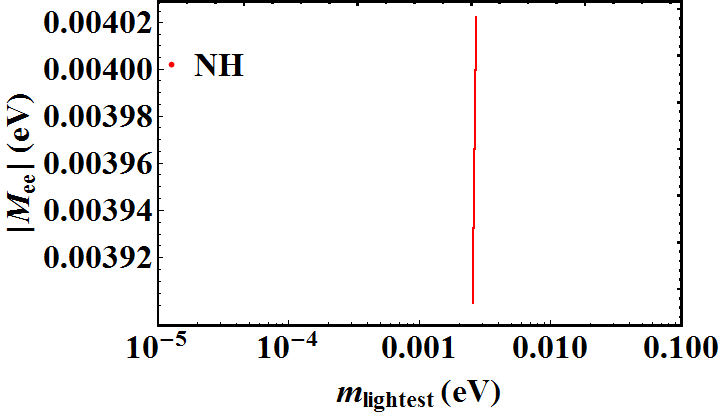} 
     \includegraphics[width=0.45\textwidth]{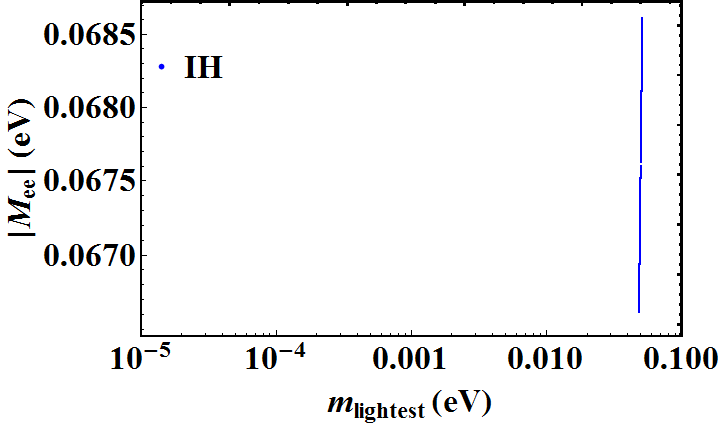} 
   \caption{Variation of effective neutrino mass with the lightest neutrino mass in the model with charged lepton correction.}
  \label{fig14} 
\end{figure*}

For the same set of allowed parameters, numerically evaluated for both the models, we also calculate the respective predictions for neutrinoless double beta decay and plot it as a function of the lightest neutrino mass. Figure \ref{fig12} shows the predictions for effective neutrino mass for both the hierarchies in the model where $\theta_{13} \neq 0$ is generated from neutrino sector itself. As expected, the inverted hierarchy predictions lie very close to the upper bound on $M_{ee}$ from KamLAND-Zen experiment \cite{kamland2}. Similarly, fig \ref{fig13} shows the predictions for effective neutrino mass $M_{ee}$ for the second model where the charged lepton mass matrix is given a correction to generate non-zero $\theta_{13}$. Due to very few number of allowed points in this case, the predicted values of $M_{ee}$ are seen as a dot for both the hierarchies. This is also due to the fact the neutrino mass sum rule \eqref{eq:sumrule} is valid in this case which restricts the lightest neutrino mass to a small range of values. As can be seen from figure \ref{fig13}, the latest KamLAND-Zen data already disfavour this case for inverted hierarchy. If we zoom the points near the two dots in figure \ref{fig13}, they look like the points shown in figure \ref{fig14}. It is interesting to note that in both the models, the Planck bound on the sum of absolute neutrino mass $\sum_i \lvert m_i \rvert < 0.17$ eV \cite{par15} results in an upper bound on the lightest neutrino mass as $m_{\text{lightest}} \leq 0.04939 \; \text{eV}$ for normal hierarchy, $m_{\text{lightest}} \leq 0.0414 \; \text{eV}$ for inverted hierarchy, if we use the best fit values of mass squared differences. Interestingly this bound almost coincides with the bound from the KamLAND-Zen experiment as seen from figure \ref{fig12}.
\begin{figure}
\includegraphics[width=0.75\linewidth]{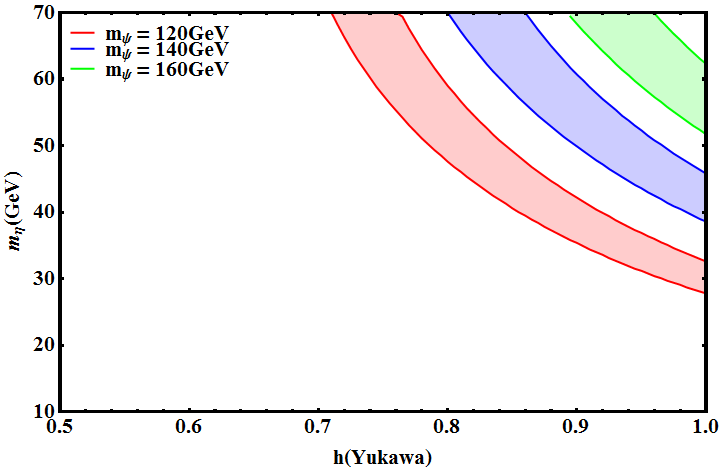}
\caption{Dark matter mass as a function of Yukawa coupling keeping the mediator mass fixed for each plots, such that the constraints on the DM relic abundance is satisfied.}
\label{fig15}
\end{figure} 

Finally we show the allowed range of dark matter mass and its couplings to leptons from the requirement of satisfying correct dark matter relic abundance criteria in figure \ref{fig15}. As expected, higher the values of mediator mass, the larger Yukawa couplings are needed to give rise to the correct relic abundance. Such large Yukawa couplings and smaller mediator masses favourable from lepton portal limit of DM will make the charged lepton correction case more favourable. This is because, one needs suppressed Yukawa couplings or large mediator mass in order to generate sub-eV corrections to light neutrino mass, than generating sub-GeV corrections to the charged lepton mass matrix.

\section{Conclusion}
\label{sec:level7}
We have studied a TeV scale inverse seesaw model based on $S_4$ flavour symmetry which can naturally generate correct light neutrino masses with Tri-Bimaximal type mixing at leading order. The model also predicts a neutrino mass sum rule that can further predict the value of the lightest neutrino mass, that can be tested at experiments like neutrinoless double beta decay. Since TBM mixing has already been ruled out by the latest neutrino oscillation data, we consider two possible ways of generating non-zero $\theta_{13}$ which automatically take dark matter into account. The idea is based on the scotogenic mechanism of neutrino mass generation, where neutrino mass arises at one loop level with DM particles going inside the loop. We first give such a one loop correction to the leading order light neutrino mass matrix and numerically evaluate the model parameters from the requirement of satisfying the correct neutrino data. This however, disturbs the mass sum rule prediction of the original model. The dark matter candidate in such a case could either be a singlet neutral fermion or the neutral component of a scalar doublet, depending whichever is lighter. We also study the possibility of generating $\theta_{13}\neq 0$ by giving a correction to the charged lepton sector. Such a case is found to be more constrained from the requirement of satisfying the correct neutrino data. We find much narrower ranges of points in terms of light neutrino parameters which can bring the model predictions closer to the observed data. Consistency with light neutrino data also requires the right diagonalising matrix of charged lepton to have very small mixing angles. The DM candidate in this case is the neutral component of a scalar doublet.

We also study the predictions for neutrinoless double beta decay and found that the charged lepton correction case with inverted hierarchy is disfavoured by the latest KamLAND-Zen data. The predictions for effective neutrino mass in this model is very specific and confined to a tiny region around a particular value of lightest neutrino mass. This is due to the neutrino mass sum rule which forces the lightest neutrino mass to remain within a very narrow range. We also find the allowed parameter space for scalar dark matter from the requirement of producing the correct neutrino data, ignoring the Higgs portal and gauge mediated annihilations. Such lepton portal annihilations are efficient for large Yukawa couplings or smaller mediator masses. Since the same Yukawa couplings and mediator mass go into the one loop correction for both neutrino and charged lepton mass matrix, the charged lepton correction is more favourable from lepton portal scalar DM point of view. As mentioned before, this is due to the fact that large Yukawa or small mediator mass will be able to generate sub-GeV corrections to charged lepton mass matrix more naturally than generating sub-eV corrections to light neutrino mass matrix. Also, the charged lepton correction case is much more predictive, as obvious from a much narrower region of allowed parameter space compared to the model with neutrino mass correction.

 \appendix
\section{Properties of $S_4$ group} 
\label{appen1}

 $S_{4}$ is the group of permutations of four objects. It has got 24 group elements.
 There are five inequivalent irreducible representations
  of $S_{4}$, among which there are two singlets 1 and $1^{\prime}$, one doublet 2 and two triplets 3 and $3^{\prime}$.
 The representations are given as follows
 \begin{gather*}
  a,b \sim1_{1} ,
  \left(\begin{array}{c}
    a_{1}\\
    a_{2}\end{array}\right) ,  
   \left(\begin{array}{c}
     b_{1}\\
    b_{2}\end{array}\right) \sim 2 , \left(\begin{array}{c}
    a_{1}\\
    a_{2}\\
    a_{3}
   \end{array}\right),  \left(\begin{array}{c}
     b_{1} \\
    b_{2} \\
    b_{3}\end{array}\right) \sim 3 , \left(\begin{array}{c}
    a_{1}^{\prime} \\
    a_{2}^{\prime} \\
    a_{3}^{\prime}
   \end{array}\right),  \left(\begin{array}{c}
     b_{1}^{\prime} \\
    b_{2}^{\prime} \\
    b_{3}^{\prime} \end{array}\right) \sim 3^{\prime}.
 \end{gather*} 
 
 The tensor products of $S_{4}$ that has been used in the present analysis are given below (for more details see \cite{Alt10})
 \begin{equation*}
 3 \otimes 1 = 3, 3 \otimes 1^{\prime} = 3^{\prime}, 3^{\prime} \otimes 1^{\prime} = 3 ,  2 \otimes 1^{\prime}=2.\\
 \end{equation*}
 
 \begin{gather}
 (A)_{3} \otimes (B)_{3} = \left(A \cdot B \right)_{1} \oplus \left(\begin{array}{c}
    A \cdot \Sigma \cdot B\\
    A \cdot \Sigma^{*} \cdot B \end{array}\right)_{2} \oplus \left(\begin{array}{c} 
    \{A_{y}B_{z}\}\\
    \{A_{z}B_{x}\}\\
    \{A_{x}B_{y}\} \end{array}\right)_{3} 
    \oplus \left(\begin{array}{c} 
     [A_{y}B_{z}]\\
     \left[A_{z}B_{x}\right]\\
     \left[A_{x}B_{y}\right] \end{array}\right)_{3^{\prime}}.     
 \end{gather}
\begin{equation}
 \begin{gathered}
  A \cdot B = A_{x}B_{x}+ A_{y}B_{y} +  A_{z}B_{z}\\ 
  \{A_{i}B_{j}\} = A_{i}B_{j} + B_{j}A_{i}\\ 
  \left[A_{i}B_{j}\right] = A_{i}B_{j} - A_{j}B_{j}\\   
  A \cdot \Sigma \cdot B = A_{x}B_{x} + \omega A_{y}B_{y} + \omega^2 A_{z}B_{z}\\  
  A \cdot \Sigma^{*} \cdot B = A_{x}B_{x} + \omega^2 A_{y}B_{y} + \omega A_{z}B_{z}.  
 \end{gathered}
 \end{equation}
Later on for simplicity, we can replace $3 \rightarrow 3_{1}$, $3^{\prime} \rightarrow 3_{2}$, $1 \rightarrow 1_{1}$, 
 $1^{\prime} \rightarrow 1_{2}$.
 \begin{gather*}
 2 \otimes 2= 1_{1} \oplus 1_{2} \oplus 2,\\
   3_{1} \otimes 3_{1}= 1_{1} \oplus 2 \oplus 3_{1} \oplus 3_{2}.
 \end{gather*}
 The Clebsch-Gordon coefficients for $3_{1}\times 3_{1}$, used in our analysis is as follows    
 \begin{gather*}
  \left(\begin{array}{c}
         a_{1}\\
         a_{2}\\
         a_{3} \end{array}\right)_{3_{1}}\otimes \left(\begin{array}{c}
         b_{1}\\
         b_{2}\\
         b_{3} \end{array}\right)_{3_{1}} = (a_{1}b_{1}+a_{2}b_{2}+a_{3}b_{3})_{1_{1}} \oplus \left(\begin{array}{c}
         1/\sqrt{2}(a_{2}b_{2} - a_{3}b_{3})\\
         1/\sqrt{6}(-2a_{1}b_{1}+ a_{2}b_{2} + a_{3}b_{3})
          \end{array}\right)_{2} \oplus  \\
          \left(\begin{array}{c}
         a_{2}b_{3}+a_{3}b_{2}\\
         a_{1}b_{3}+a_{3}b_{1}\\
         a_{1}b_{2}+a_{2}b_{1} \end{array}\right)_{3_{1}} \oplus \left(\begin{array}{c}
         a_{3}b_{2}-a_{2}b_{3}\\
         a_{1}b_{3}-a_{3}b_{1}\\
         a_{2}b_{1}-a_{1}b_{2} \end{array}\right)_{3_{2}}.
   \end{gather*}
  \section*{Acknowledgements}
  AM acknowledges the hospitality and laboratory facilities provided by the Department of Physics, IIT Guwahati during her visit in February 2017 when some part of this work was completed. The work of MKD is partially supported by the grant no. 42-790/2013(SR) from
University Grants Commission, Government of India.
\bibliographystyle{apsrev}

\end{document}